\documentclass[]{article}

\usepackage{listings}
\usepackage[margin=0.75in]{geometry}
\usepackage{graphicx}
\usepackage{epstopdf}
\usepackage{color}
\usepackage{amsmath}
\usepackage{amssymb}
\usepackage{permute}
\usepackage{etoolbox}
\usepackage{tikz}
\usepackage[all]{xy}
\usepackage[utf8x]{inputenc}
\usepackage{import}
\usepackage{setspace}
\usepackage{authblk}
\usepackage{hyperref}
\usepackage{xfrac}
\usepackage{csquotes}
\usepackage{multicol}
\usepackage{wrapfig}
\usepackage{bm}

\newcommand\mperiod[1][\rlap]{#1{\;\;\;.}}

\title{Variability in higher order structure of noise added to weighted networks}

\author{Ann S. Blevins, Jason Z. Kim, Danielle S. Bassett}

\begin{document}

\maketitle

\section{Abstract}

From spiking activity in neuronal networks to force chains in granular materials, the behavior of many real-world systems depends on a network of both strong and weak interactions. These interactions give rise to complex and higher-order system behaviors, and are encoded using data as the network's edges. However, distinguishing between true weak edges and low-weight edges caused by noise remains a challenge. We address this problem by examining the higher-order structure of noisy, weak edges added to model networks. We find that the structure of low-weight, noisy edges varies according to the topology of the model network to which it is added. By investigating this variation more closely, we see that at least three qualitative classes of noise structure emerge. Furthermore, we observe that the structure of noisy edges contains enough model-specific information to classify the model networks with moderate accuracy. Finally, we offer network generation rules that can drive different types of structure in added noisy edges. Our results demonstrate that noise does not present as a monolithic nuisance, but rather as a nuanced, topology-dependent, and even useful entity in characterizing higher-order network interactions. Hence, we provide an alternate approach to noise management by embracing its role in such interactions.

\newpage

\twocolumn

\section{Introduction}

\begin{figure*}
    \centering
    \includegraphics[width=\linewidth]{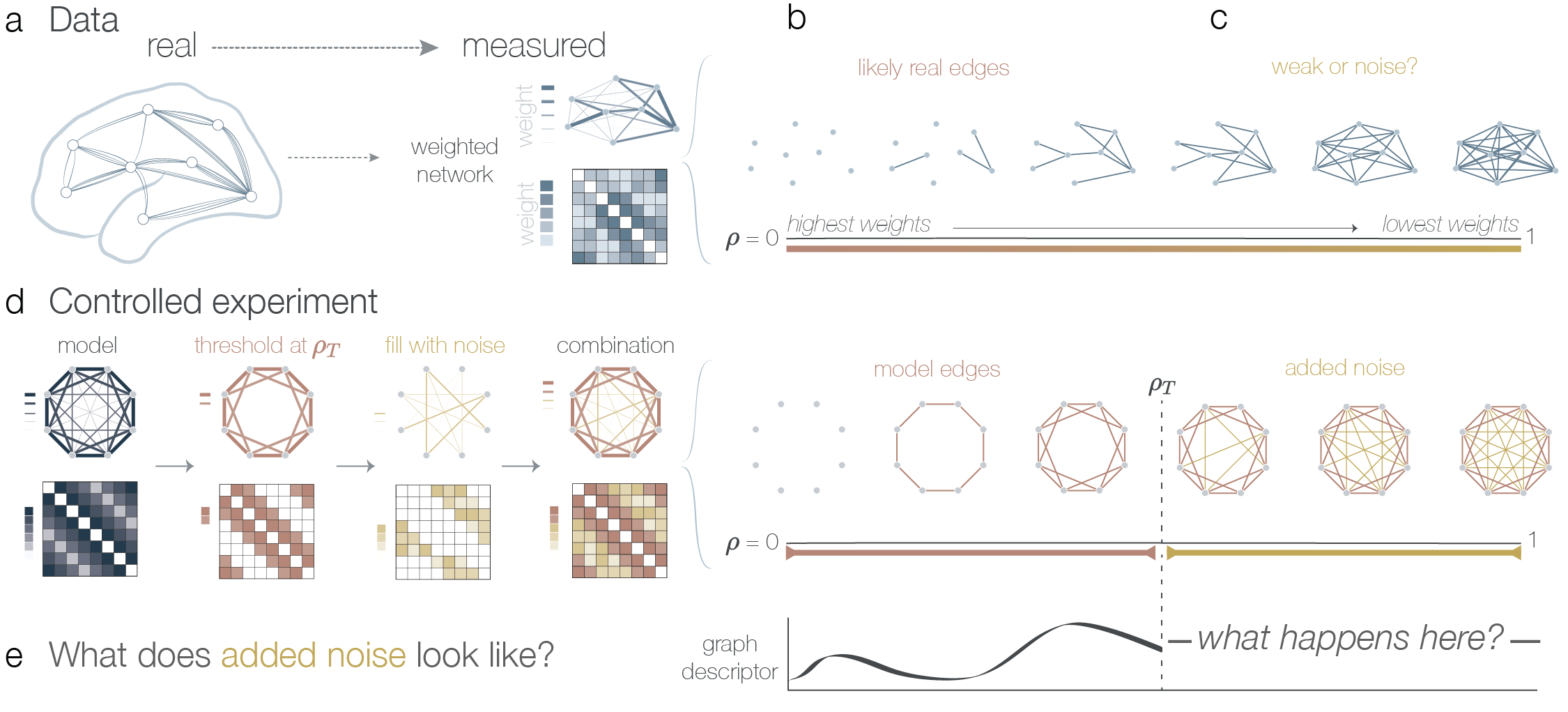}
    \caption{\textbf{Our experiments query the structure of noise added to model networks.} \emph{(a)} A schematic of real data from a structural brain network. Such data are often assumed to consist of strong and likely real edges \emph{(b)}, along with weak edges \emph{(c)} that could arise simply due to noise in the system or in the measurements. Edge density denoted by $\rho$. \emph{(d)} In our experiment, we begin with a model network (navy) that has been thresholded to a specific edge density $\rho_T$ (rose), and fill the thresholded edges with weaker, noisy edges (gold) to form a combined network (rose-gold). \emph{(e)} Then we will measure the structure of the noisy edges (gold).}
    \label{fig:intro}
\end{figure*}

In weighted network analyses, it is incredibly difficult to distinguish between weak edges that correspond to significant system features and weak edges that are simply false positives or noise. Previous research in neuroscience \cite{santarnecchi2014efficiency,goulas2015strength,bassett2012altered}, social networks \cite{granovetter1973strength,friedkin1980test}, and molecular biology \cite{ma2012discovering} has demonstrated the importance of weak, real edges, to the function of a network. However, weighted networks collected from data are also often plagued by noise either in the form of imprecise edge weights or spurious connections. Unfortunately, such noise often affects the signal-to-noise ratio of weak edges more than strong edges, thus complicating our understanding of weak edge network structure.

How does one deal with unwanted noise in estimates of network structure in real systems? Current methods often attempt to remove noisy edges via thresholding \cite{bordier2017graph,de2013estimating, serrano2009extracting,alexander2010disrupted,wijk2010comparing} by taking into account edge weight, density, group similarity, or network measures \cite{bordier2017graph}. However, thresholding presents a challenge as one may remove too many real edges or keep too many spurious edges \cite{zalesky2016connectome}. Alternatively, many theoretical studies focus on structural properties of noisy edges independent of and isolated from any data; most commonly, efforts in this space study random graphs. One can predict graph properties of random networks \cite{wigner1958distribution,ramezanpour2003generating,erdos1959random,chung2016decomposition}, which can be useful in distinguishing real networks from random graphs. However, noisy edges in real networks do not exist in isolation, but are intertwined with real edges. Consequently, neither thresholding nor studying noise in isolation is without flaws when distinguishing the structure of real weak edges from that of noisy weak edges. Instead of isolating noise, can we understand its role in higher-order network structure to mitigate, and potentially even take advantage of, noisy edges?

Here we address the above questions by welcoming noise into our experiments and descriptions of network structure (Fig.~\ref{fig:intro}). Specifically, we ask what (if anything) can noise added to a real network tell us about the underlying real network structure? If the added noise has the same structure regardless of the topology of the real network, then the answer is nothing. The advantage of this scenario is that we could potentially identify an appropriate threshold for our data fairly easily and in a data-driven manner. Instead, if the structure of the added noise varies based on the topology of the real network, then the noise may carry information about the real weighted network. Indeed, given that a binary graph is defined by a list of either its edges or its non-edges, it is possible that the noise filling the empty edges of a real weighted network holds information related to the topology of the real edges.

To explore this possibility, we ask whether the higher order structure of noise varies across model network topologies in the controlled context of adding noise to model networks. We tested twelve model networks that spanned different node strength distributions, reliance on distances, and more. Using the network models tested, we identify at least three common profiles of added noise structure as assessed by persistent homology. Notably, we find that these patterns correspond to the topologies of the model networks to which the noise was added. We find that enough information exists within the structure of added noise alone to reasonably distinguish between network models, and that this information stems both from features persisting from the model network and those formed by noisy edges. Furthermore we provide generative rules that explain the different patterns of added noise structure. Finally we remark on the ability of noise to create misleading structure within weak network edges and discuss the implications for data analysis.

\section{Results}
\subsection{Experimental setup}

To better understand the effect of noise added to networks, we precisely measure the changes in network structure that arise from noisy, weak edges, and compare them to an expected structure. We begin with a weighted, completely connected model network generated from a specific set of rules, which has a predictable structure (Fig.~\ref{fig:intro}d, navy). Next, we create a space for weak, noisy edges by thresholding this model network at a chosen edge density $\rho_T$ to keep only those strongest $\rho_T$ fraction of edges (rose). We then create the added noise network (gold) by assigning a random weight to any edge not included in the $\rho_T$-thresholded weighted network. All edges in the added noise network have edge weight less than any edge in the $\rho_T$-thresholded network. Finally, we combine the $\rho_T$-thresholded network and the added noise network to yield a combined weighted network (rose and gold), which now has a precise cutoff below which all edge weights are randomly chosen. Said another way, if we begin with an empty network and add edges to this graph in order of decreasing edge weights from the combined weighted network, the first $|E|\rho_T$ edges will come from modeled edges while the latter $1-|E|\rho_T$ edges will come from noisy edges only. Then if we measure network structure along this expanded version of the combination network (Fig.~\ref{fig:intro}d, right), we will be able to distinguish between structure attributed to the model edges from that created by the added noise.

\subsection{Mathematical framework}

\begin{figure*}
    \centering
    \includegraphics[width=\linewidth]{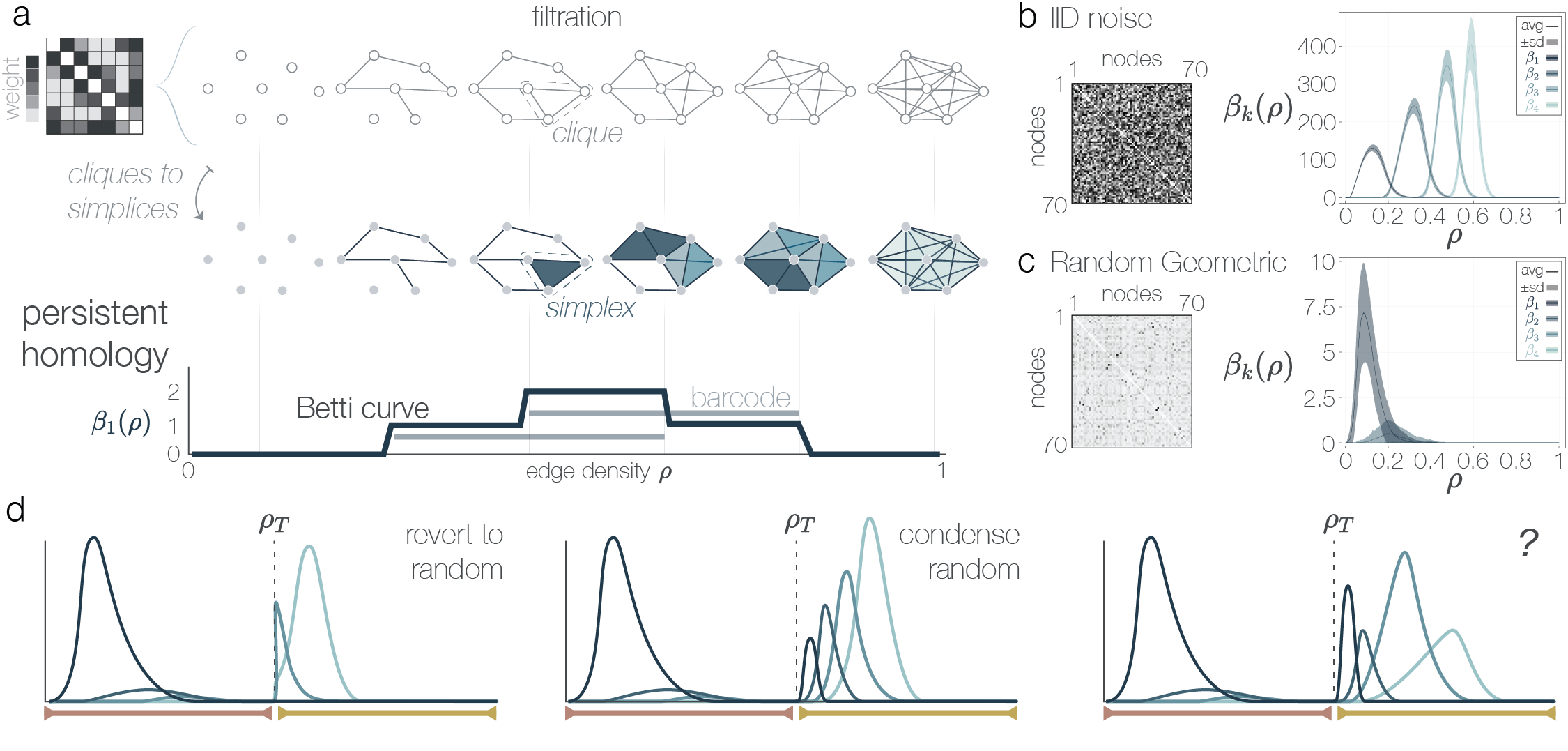}
    \caption{\textbf{Through the lens of persistent homology, we imagine three possibilities for the structure of noise added to model networks.} \textit{(a)} From a weighted network, we construct a graph filtration (top) by adding edges to an empty graph one at a time in order of decreasing edge weight. From this filtration we create a sequence of clique complexes (middle) on which we compute the persistent homology (bottom). We show the barcode (horizontal lines) and Betti curve $\beta_1(\rho)$ for the resulting persistent homology. \textit{(b, c)} An example matrix (left) and Betti curves (right) of the IID noise \textit{(b)} and random geometric \textit{(c)} network models. Solid lines indicate the average over 500 replicates and shaded ares correspond to $\pm 1$ standard deviation. \textit{(d)} Stylized possibilities for the structure of added noise as perceived by the Betti curves: a reversion to IID noise Betti curves (left), a condensed collection of all four Betti curves following an IID noise pattern (middle), or a Betti curve pattern completely unlike that of IID noise (right).}
    \label{fig:setup}
\end{figure*}

Real data is often characterized by three features: weighted relationships between nodes, higher order interactions, and topological constraints such as wiring distance. Given these features, we use persistent homology to study weighted network structure. Persistent homology \cite{carlsson2009topology,zomorodian2005computing,otter2017roadmap} records the longevity of topological cavities that form and collapse throughout the graph filtration $$G_0 \subseteq G_1 \dots \subseteq G_{|E|}\mperiod$$ This filtration is the formalization of the expanded view of a weighted network discussed in Fig.~\ref{fig:intro}, in which $G_i$ is the binary graph containing the $i$ strongest edges in the weighted network (see Fig.~\ref{fig:setup}a, Methods, and \cite{sizemore2016classification,giusti2015clique,petri2013topological, horak2009persistent}). This persistent homology approach has been previously used to identify differences in cognition across individuals based on resting state functional connectivity \cite{anderson2018topological}, to find percolation properties of porous materials \cite{robins2016percolating}, to differentiate neuron morphologies \cite{kanari2020trees},  and to understand many other real-world systems \cite{curto2017can,hess2020topological,patania2017shape}. We note that in our setup, only the rank order of edges induced by the original edge weights are preserved, so that the specific edge weights or their generating distribution does not affect the outcome. For this work, we compute the persistent homology in several dimensions: dimension 1 (gaps surrounded by edges), 2 (voids surrounded by filled triangles), 3 (voids surrounded by collections of tetrahedra), and 4 (a higher dimensional analog). A filtration of graphs can be translated into a filtration of higher-order complexes called simplicial complexes on which we can compute persistent homology, by assigning a clique of $k+1$ nodes to a $k$-simplex (Fig.~\ref{fig:setup}a, see Methods for more details).

The persistent homology of a weighted network is a collection of half-open intervals $[b_l,d_l)$ called the barcode that denote the birth $b_l$ and death $d_l$ of the $l^{th}$ persistent cavity in dimension $k$. We visualize this output as either the barcodes itself or a Betti curve summarization (Fig.~\ref{fig:setup}a, bottom). In the barcode visualization, each bar corresponds to a persistent cavity and extends from the bar's birth to its death. In the Betti curve plot, $\beta_k(\rho)$ counts the number of persistent cavities alive at edge density $\rho$. Unlike many other graph metrics, persistent homology incorporates the strong and weak interactions in a holistic manner by considering how the entire filtration fits together, which both generates a unique perspective on network structure similarity \cite{sizemore2016classification} and allows us to more precisely understand the interplay between edges corresponding to real data and edges added randomly.

As motivating examples, we present the persistent homology of two random systems \cite{kahle2009topology,kahle2011random}. For both systems, we summarize their persistent homology using Betti curves as shown in Fig.~\ref{fig:setup}. In the first system, we create a noisy network by assigning to every edge a weight sampled uniformly at random from $(0,1)$. We denote this model by `IID noise' for the duration of the paper. For the IID noise model, we observe the characteristic increase of the Betti curve peaks with increasing dimension \cite{kahle2009topology,kahle2012sharp} (Fig.~\ref{fig:setup}b). Next, we consider the Betti curves of another well-studied model: the random geometric complex \cite{kahle2011random}. In this second model system, we choose points uniformly at random from the unit cube and weight edges as the inverse Euclidean distance between each pair of points (see Methods). Note that the random geometric Betti curves (Fig.~\ref{fig:setup}c) are an order of magnitude smaller than those of the IID noise model, and that the peaks decrease with increasing dimension.

Combining persistent homology with the experimental approach set up in Fig.~\ref{fig:intro}, we imagine three possibilities for the impact of added noise on the Betti curves of model networks, illustrated with stylized Betti curves in Fig.~\ref{fig:setup}d. First, after $\rho_T$ the Betti curves could quickly revert to the expected IID noise pattern at that edge density (Fig.~\ref{fig:setup}d, left). Then we would see that the noise section of the Betti curves looks similar to a copy-and-pasted version of that same section of the IID noise Betti curves. Second, we could imagine that if the weighted network model has one densely connected community, there will be so much empty space that randomly adding edges will create a smaller version of the IID noise network (Fig.~\ref{fig:setup}d, middle). We might expect this scenario to produce the entire sequence (non-zero $\beta_k$ for $k = 1, \dots, 4$) of increasing Betti curves all condensed after $\rho_T$. Third, perhaps neither of the above is correct, and in fact the added noise section of the filtration may show no resemblance to the IID noise Betti curves (Fig.~\ref{fig:setup}b, right). Which of these three possibilities actually occurs?

\subsection{Noise structure varies across model networks} \label{sec:variation}

\begin{figure*}
    \centering
    \includegraphics[width=\textwidth]{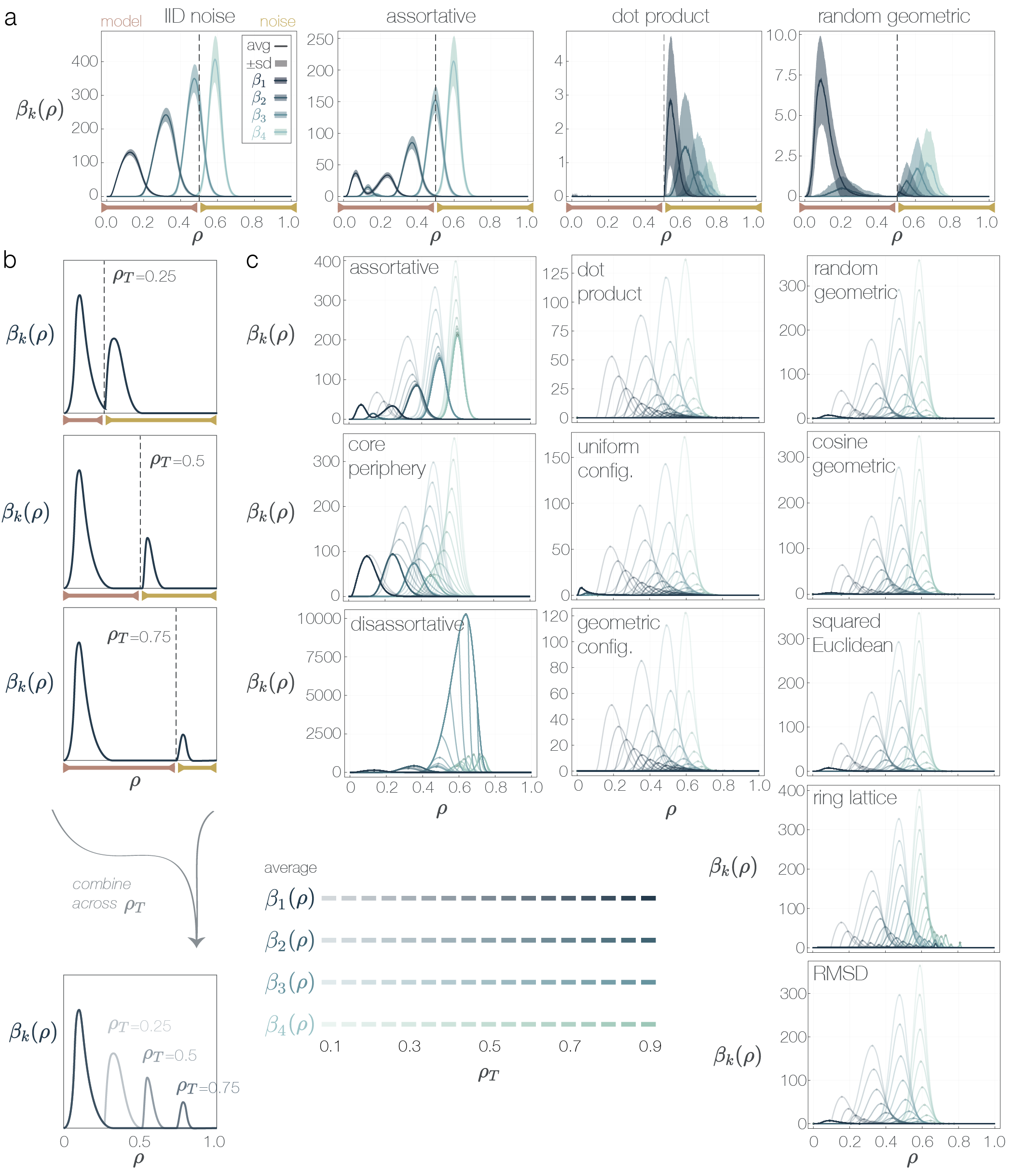}
    \caption{\textbf{Betti curves of added noise vary based on the underlying network model.} \textit{(a)} Examples of network models that support varying structure of added noise. Solid lines indicate mean Betti curve and shaded region corresponds to $\pm 1$ standard deviation. See Fig.~\ref{sfig:bettis_05} for all models. \textit{(b)} Repeating across values of $\rho_T$, we create a summarized visualization of the Betti curves. \textit{(c)} Mean Betti curves for each of the 11 tested network models (IID noise not included). Betti curve opacity indicates $\rho_T$. Betti curve peaks are marked with a dot to emphasize the scaling and shifting of Betti curves as $\rho_T$ varies.}
    \label{fig:1}
\end{figure*}

We test which of the three scenarios described above occurs for 12 graph models (including IID noise, see Methods) and 17 values of $\rho_T \in [0.1, 0.9]$. We chose the following graph models to span distinct generative rules, topological characteristics, and overall structure; IID noise, assortative, core periphery, cosine geometric, disassortative, (discrete) uniform configuration, dot product, geometric configuration, random geometric, ring lattice, root mean squared deviation (RMSD), and squared Euclidean (see Methods). Among these model networks, we observe examples of each of the three possible scenarios (Fig.~\ref{fig:1}a). Specifically, we observe that the structure of the added noise network atop the assortative model with $\rho_T = 0.5$ is qualitatively similar to that of the IID noise model. The dot product model supports added noise that does not follow any expected pattern, and instead has a decreasing pattern of peaks with increasing dimension. Finally, we observe that the added noise atop the random geometric model produces all four dimensions of persistent homology in an increasing pattern of peaks after $\rho_T=0.5$. We include results for all models in Fig.~\ref{sfig:bettis_05}.

Repeating the experiment across all 17 values of $\rho_T$ (Fig.~\ref{fig:1}b), we observe three qualitative classes of added noise structure (Fig.~\ref{fig:1}c) within the non-IID noise models. We will refer to the three qualitative classes as the \textit{random reversion}, \textit{coned}, and \textit{random condensed} classes. First, the Betti curves of the added noise networks on the assortative, core periphery, and to a lesser extent the disassortative models mirror those of the IID noise model but scaled and sometimes slightly shifted; these models comprise the \textit{random reversion} class, named for the return of the Betti curves to those of the random IID noise model. Second, noise added to the configuration and the dot product models generate Betti curve peaks that decrease with increasing dimension and dramatically shift rightwards as $\rho_T$ increases. We will refer to this second set of models as the \textit{coned} class following Ref. \cite{sizemore2016classification}. Third, the distance-based models (random geometric, cosine geometric, ring lattice, squared Euclidean, and RMSD) constitute the \textit{random condensed} class. Here, the added noise produces an increasingly compressed collection of IID noise-like Betti curves in all four dimensions as $\rho_T$ increases. In summary, we found that the structure of noise added to model weighted networks varies across network models and values of $\rho_T$.

\subsection{Noise added to networks can distinguish many network models.} \label{sec:classification}

We demonstrated that the structure of noise varies based on the model to which it is added. But does the structure of added noise vary \emph{enough} that it can accurately classify the network models? In the binary case, there is as much information about a binary graph from knowing all the edges that exist as there is from knowing which edges do not exist. Our experiment could be interpreted as a weighted extension of this idea, in which we know weights of model network edges, and then all the open space (edges with weight 0 after thresholding at $\rho_T$) are filled by random weights. The randomly-weighted edges and the model-weighted edges together form a complete (weighted) graph. Therefore, despite the fact that edge weights are chosen at random for the added noise, we expect that at $\rho_T$ near 0.5 the added noise persistent homology will be enough to classify the model networks. 

\begin{figure*}
	\centering
	\includegraphics[width =\linewidth]{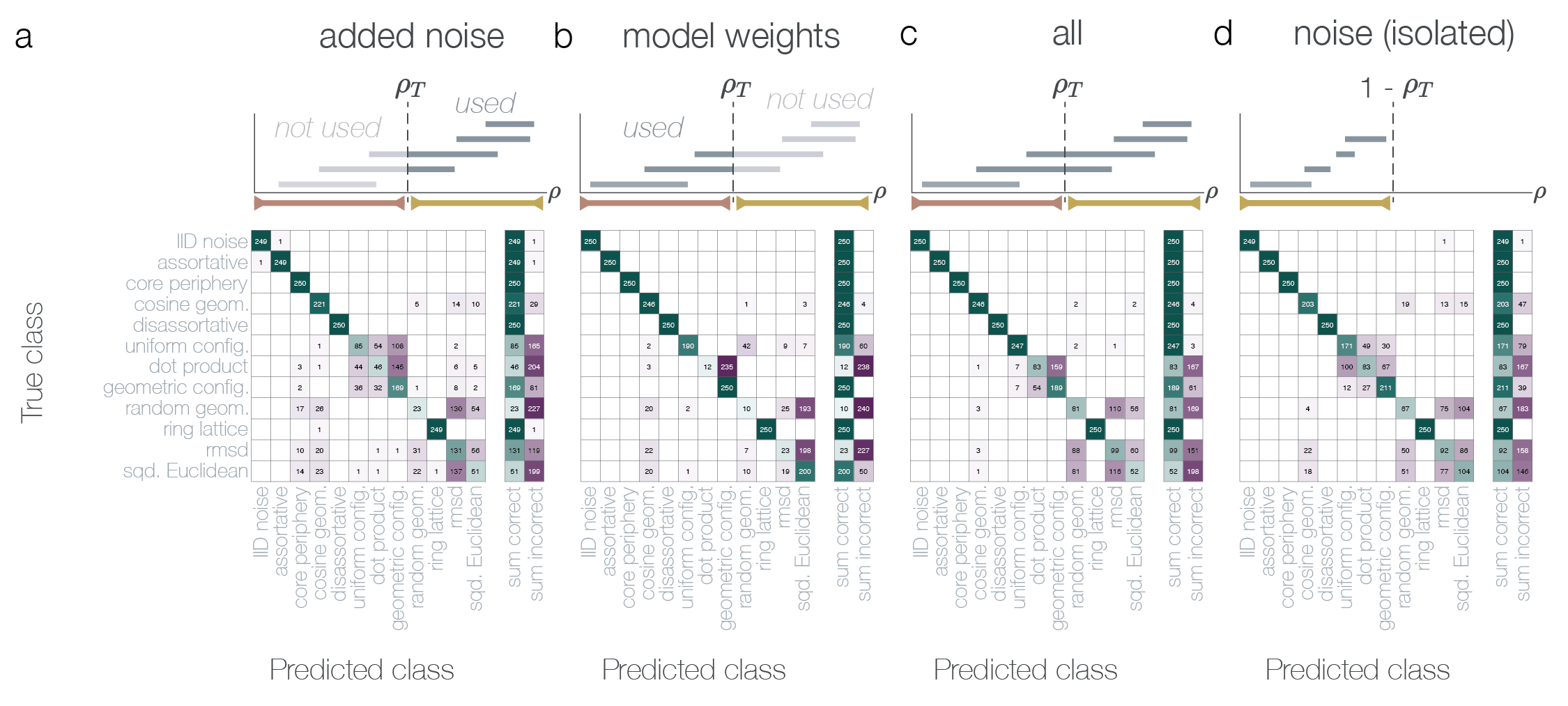}
	\caption{\textbf{Classification based on the added noise persistent homology can distinguish between network models.} Confusion matrices showing the results of the classifications based on \textit{(a)} added noise barcodes, \textit{(b)} the barcode from model weights only, \textit{(c)} the entire barcode, and \textit{(d)} barcodes generated from the isolated added noise networks. Matrix entries show the number of corresponding true models that were classified as the corresponding predicted model, out of 250 (see Methods).}
	\label{fig:2}
\end{figure*}

To test these ideas, we classify the networks using a Gaussian mixture model. We use three features derived from the persistent homology in dimension $k$, for $k=1,\dots, 4$ (12 features total). Described in detail in the Methods section, these three features quantify the total amount of persistent homology (sum of persistent lifetimes), and the lifetime sum weighted by either the birth or death time of persistent cavities \cite{adcock2012ring,sizemore2016classification}. We can subset the barcode based on the desired classification (for example, using only the barcode from the added noise portion of the filtration), and we perform the classification procedure for each value of $\rho_T$; see confusion matrices for $\rho_T = 0.5$ in Fig.~\ref{fig:2}, classification accuracy in Fig.~\ref{sfig:classification_accuracy}, and Section \ref{sclassification} for all results.

We find that at $\rho_T=0.5$ the classification performs almost similarly when using features from the added noise portion of the filtration (Fig.~\ref{fig:2}a, accuracy$\approx0.658$) over the model network portion (Fig.~\ref{fig:2}b, accuracy$\approx0.722$). Notably, for particular network models such as the RMSD and dot product models, the classification using added noise features outperformed the classification with the model features. Other models such as the squared Euclidean and uniform configuration models were better classified using model weights than the added noise portion of the barcode. Finally, when we classify using the entirety of the barcodes (Fig.~\ref{fig:2}c), we find that the prediction accuracy is at least as good or improved over both the added noise and model weights section for eight of the twelve models. Indeed, using all barcodes results in the highest accuracy for $0.2 \leq \rho_T \leq 0.6$ (Fig.~\ref{sfig:classification_accuracy}, left). This general accuracy improvement suggests that the added noise and model portions of the barcode contain non-overlapping information.

Surprisingly, at $\rho_T=0.5$ we also find a comparably good classification when using features from a graph that contains \textit{only} the noisy edges, with all model edges set to 0. As we have previously discussed, the added noise edges on any weighted network is itself a weighted graph. The above classification experiment used the added noise structure that was dependent -- that is -- existed atop a weighted network, such that the added noise existed \textit{within the context of} the model network. That approach leaves open the question of whether this added noise graph as a standalone weighted network (independent of any model edges already added) contains the same amount of information as the added noise network (noise in the context of the network model)? To address this question, we extracted the added noise network from each graph model for all values of $\rho_T$ and computed their persistent homology. We show the Betti curves across values of $\rho_T$ for these isolated noise networks in Fig.~\ref{sfig:noiseonly}. We find that classifying the topological features from the added noise independent of model edges (noise (isolated)) does a fine job (accuracy $\approx0.719$, see Fig.~\ref{sfig:classification_accuracy}) in comparison to the added noise in-context features (Fig.~\ref{fig:2}d). In fact, for 10 of the 12 models the classification matches or outperforms that using the added noise in context, and for three models it outperforms the classification using model weights only. These results indicate that randomly adding edges drawn from a specific organization (here, the complement of the model network at $\rho_T$) contains an impressive amount of distinguishing power.

Finally, as a byproduct of studying the isolated added noise networks (Fig.~\ref{sfig:noiseonly}) we learn that the disassortative and assortative models are, in a way, weighted complements of each other. Indeed the space left by the thresholded disassortative network at particular values of $\rho_T$ exists within four large groups of nodes, so that the weighted random complement is indeed structured as four communities. We see evidence of this type of community structure in the Betti curves, particularly in the two peaks observed $\beta_1$ and $\beta_2$ \cite{sizemore2016classification} of the isolated noise networks (Fig.~\ref{sfig:noiseonly}). Taken together, we learn that the structure of noise added to model networks contains information not seen in the structure of the model network itself. In the particular case of model networks that produce little persistent homology, the structure of their empty edges as seen by the added noise persistent homology can improve classification.

\subsubsection{Determining the source of information contained within added noise topology} 

Given the ability of the added noise persistent homology features to reasonably classify the underlying network models, we next aim to clarify more precisely the reason for this result. We can label persistent cavities (bars in the barcode) that exist during the added noise section as one of two types: the first a so-called \textit{noise-exclusive} persistent cavity whose birth and death time are both within the added noise portion of the filtration ($b,d>\rho_T$), and a \textit{crossover} cavity that was born within the model section ($b<\rho_T$) but dies within the added noise section ($d> \rho_T$, see Methods). We expect that, for those graph models that have them, the crossover bars should hold the majority of the classification information since they are formed with model weight edges. After classification on these slices of the barcodes, we find that indeed the crossover bars can be used to classify the five models that consistently produce crossover bars nearly perfectly (Fig.~\ref{sfig:classification_crossover}). Surprisingly, the classification using the noise-exclusive portion (Fig.~\ref{sfig:classification_blues}) is nearly identical to that of using the added noise portion of the barcode (Fig.~\ref{fig:2}a) for all models. Together our results suggest that using only those persistent features generated within the added noise section of the barcode are sufficient for moderate accuracy in classifying this set of model networks. 

Finally, we ask how much information is contained in simply the binary network at $\rho_T$ by comparing the persistent homology from the added noise to that from a randomized model weights experiment. In this experiment, $G_0$ is empty, the binary graph at $\rho_T$ is the same model network at $\rho_T$ as before, but the ordering of model edges added has been randomized (Fig.~\ref{sfig:randomized}). We find that the classification run on the persistent homology features of these randomized model weights (Fig.~\ref{sfig:randomized}, \ref{sfig:classification_accuracy}) performs similarly to that of the added noise features when $\rho_T=0.5$. In addition to the classification results using added noise or isolated noise, these findings strengthen the intuition that the binary network at $\rho_T$ constrains the possible persistent homology outcomes generated by randomly adding edges.

In summary, our classification experiments suggest that the distinguishing information of the added noise structure comes from both crossover and noise exclusive bars. The randomized model network experiment could be viewed as a step-wise random graph process in which exactly one interior graph $G_i$, $0<i<|E|$ is fixed. Therefore results from the randomized model networks experiments additionally suggest that even having one graph predetermined between $G_0$ and $G_{|E|}$ can greatly alter the topology of the filtration between $G_0$, $G_i$ and $G_i$, $G_{|E|}$.

\subsection{Drivers of noise profiles} \label{sec:drivers}

Next, we ask how one might obtain each of the three main noise profiles observed in Fig.~\ref{fig:1}a,c. The first \textit{random reversion} profile, in which the added noise section $(\rho>\rho_T)$ is similar to an IID noise Betti curve copied at that edge density (Fig.~\ref{fig:1}a), can be replicated using matrix blocks that are themselves created through a random process \cite{aicher2013adapting}. The assortative model shows this trend for the largest range of $\rho_T$, since it has four blocks of highly weighted edges and the rest are weak but randomly weighted edges (Fig.~\ref{sfig:graphinfo_1}). The core-periphery model shows the same pattern but for a smaller range of $\rho_T$, because it has seven blocks of highly weighted edges and thus its natural break between the high and low weighted edges occurs at a larger edge density than for the assortative model. The disassortative model only has four low-weight blocks, and we observe that at very few values of $\rho_T$ it supports added noise that also produces this similar Betti curve pattern.  

\begin{figure*}
    \centering
    \includegraphics[width=\linewidth]{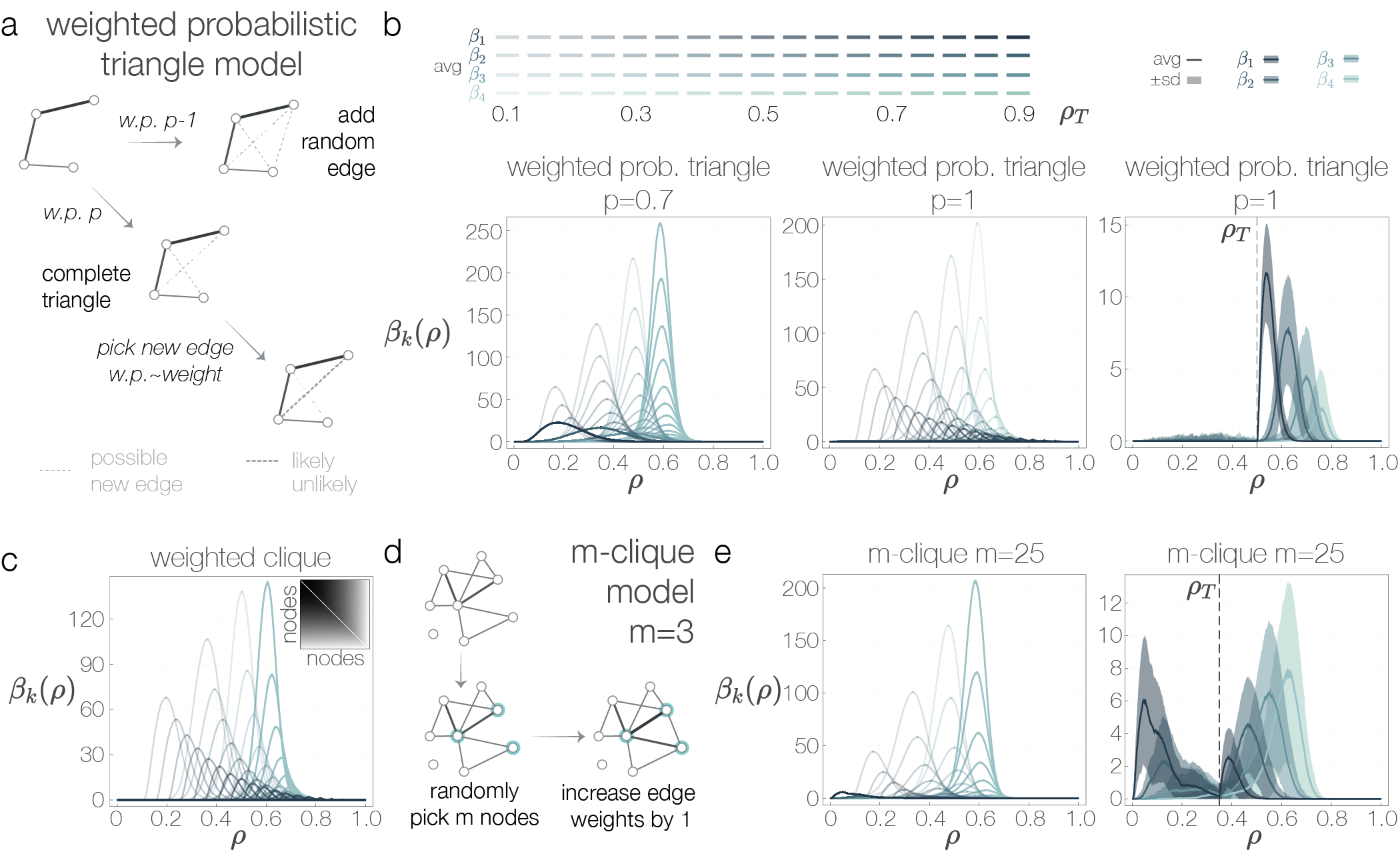}
    \caption{\textbf{Cliques and triangles support different patterns of added noise Betti curves.} \textit{(a)} The weighted probabilistic triangle model proceeds by adding edges either at random or to complete an open triangle. The likelihood of completing a specific open triangle is based on the weight of its edges. \textit{(b)} Betti curves at varying values of $\rho_T$ for the weighted probability triangle with $p=0.7$ (left) and $p=1$ (middle). Average (solid lines) and $\pm$ standard deviation (filled area) for Betti curves of the weighted probabilistic triangle model with $p=1$, $\rho_T=0.5$ (right). \textit{(c)} Betti curves at varying values of $\rho_T$ for the clique graph. Inset shows the adjacency matrix. \textit{(d)} The random $m$-clique model creates a weighted network by selecting $m$ nodes at random, increasing all edge weights between these nodes by 1, and repeating the process. \textit{(e)}. Betti curves across all values of $\rho_T$ (left) and for $\rho_T=0.35$ (right) for the $m$-clique model with $m=25$. Legends for all plots are the same as for those in panel \textit{(b)}.}
    \label{fig:3}
\end{figure*}

For the \textit{random condensed} and \textit{coned} profiles, we first investigated how the propensity to fill triangles would contribute to the added noise profile. Both the distance-based and coned network models have a strong tendency to form triangles, either by the influence of the triangle inequality in the former or the large clique size in the latter \cite{sizemore2016classification}. We create a random model network weighted so that at each step in the filtration, the next edge added completes an open triangle (three nodes connected by two edges) with probability $p$ and connects a randomly selected pair of non-adjacent nodes with probability $1-p$ (or if no open triangles exist). If the new edge will complete a triangle, the open triangle is chosen with probability weighted by the product of the open triangle edge weights (see Methods). This rule states that it is more likely for the new edge to form a triangle with two edges that were added early in the filtration than late in the filtration. We call this model the weighted probabilistic triangle model (Fig.~\ref{fig:3}a). In Fig.~\ref{fig:3}b we show the resulting Betti curves across values of $\rho_T$ for $p=0.7$ and $1$. We observe that the parameter $p$ allows us to interpolate between the IID noise network ($p=0$) and a model that supports an added noise profile of the \textit{coned} type (decreasing Betti peaks, $~0.85<p\leq 1$, see Fig.~\ref{sfig:triangle}). Taken to the most extreme, we create a weighted clique graph in which at each step in the filtration, the newest edge contributes to building one growing clique. This weighted clique model additionally shows an added noise profile similar to the weighted probabilistic triangle model with $p>0.85$ (Fig.~\ref{fig:3}c). Indeed, as discussed in the supplement, upper and lower bounds on the Betti numbers for this weighted clique network can be derived in a similar fashion to those from the IID noise graph (see Section \ref{sec:adding_noise_to_clique}).

What additional process or constraint underlies the distance-based network models that is not captured by the above weighted probabilistic triangle model? Although distance-based networks do indeed fill triangles quickly, the alternative to completing a triangle in an embedded network is far from adding a random edge anywhere in the network, as is the case in the above weighted probabilistic triangle model. Instead, in embedded networks we often see multiple pockets of clustered nodes arise and eventually connect. We aimed to capture this process at a basic, non-embedded level, by creating a model that constructs a weighted network by adding pockets of densely connected nodes to the network; we call the model the $m$-clique model (Fig.~\ref{fig:3}d). In this model, we choose a random set of $m$ nodes, increase all edge weights between these $m$ nodes by 1, and repeat this process until a desired network density is reached (see Methods). We record the persistent homology of these models and their added noise for varying values of $\rho_T$ and $m$ in Fig.~\ref{sfig:clique}, and we show the $m=25$ case in Fig.~\ref{fig:3}e. We find that for parameter values near $m=25$, the added noise Betti peaks show an increasing pattern similar to that seen with the distance-based models. We note that the random $m$-clique model does not fully capture the extent to which the Betti curve peaks shift rightwards with increasing $\rho_T$ in the distance-based models, suggesting that adding random $m$-cliques alone is not sufficient to completely recreate the observed phenomenon.

In sum, through the above generative graph models we have determined processes by which we can drive the structure of added noise towards any of the three observed profiles. 

\subsection{Misinterpretations caused by added noise}\label{sec:overlap}

We close with a more realistic example of added noise on networks, and highlight situations in which added noise may erroneously suggest non-existent network features. Above we considered the situation in which there exists a sharp, binary distinction between the model network and added noise sections of the filtration. The resulting Betti curves often show an obvious point at which the trends change drastically. Though nicer to study numerically, this situation is unlikely in real data. It is more likely that as we move along the filtration, adding stronger then weaker edges, the proportion of noisy edges increases until at some point the last real edge has been added and all the later (weaker) edges are noise.

\begin{figure*}
    \centering
    \includegraphics[width=\linewidth]{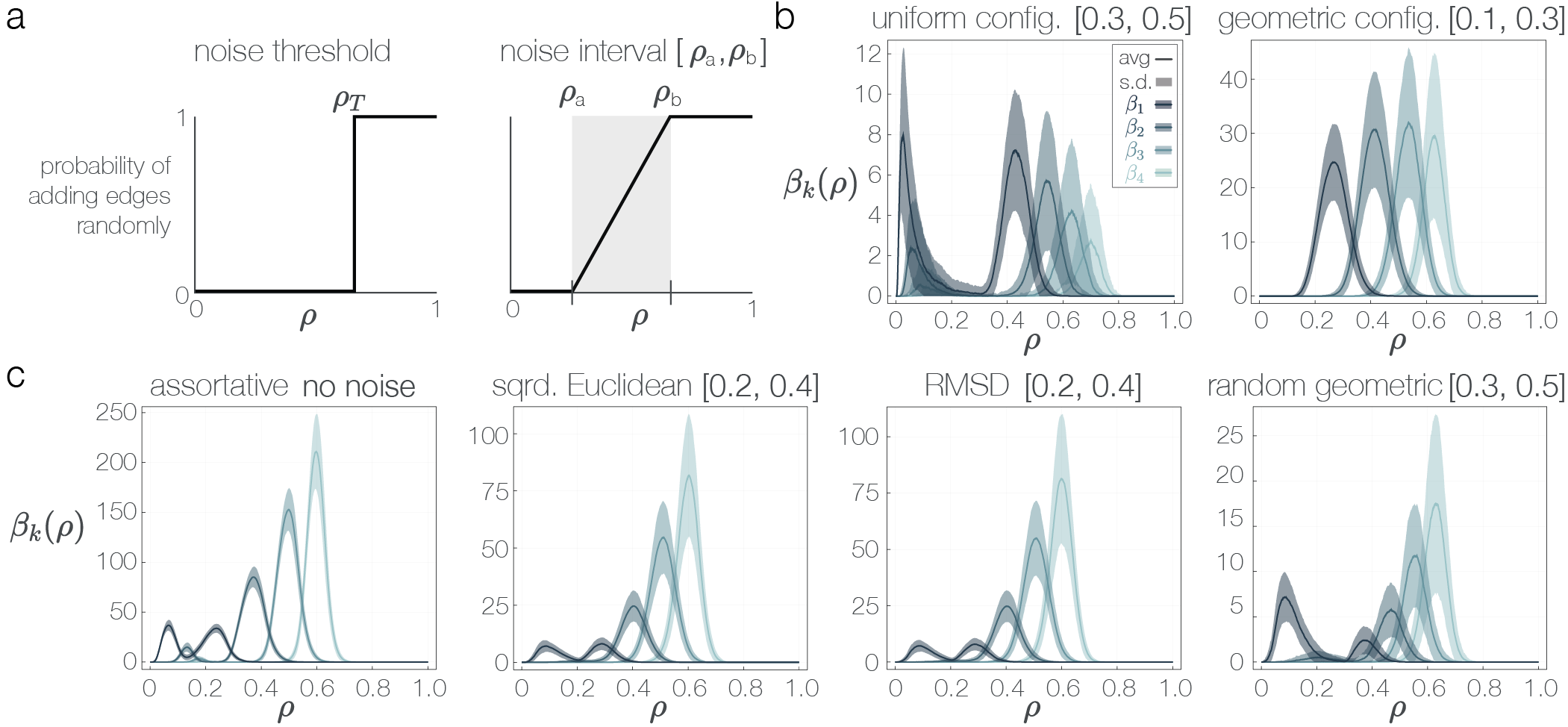}
    \caption{\textbf{Noise overlapping real networks can complicate interpretations of network structure.} \textit{(a)} Previous experiments used a sharp threshold $\rho_T$ to separate model from noise edges (left), but a more realistic scenario is to have an increasing likelihood of noisy edges over some interval $[\rho_a,\rho_b]$ (right). Betti curves from the uniform configuration model with noise interval $[0.3, 0.5]$ (left) and the geometric configuration model with noise interval $[0.1,0.3]$ (right). \textit{(c)} Betti curves for the (left to right) assortative model with no noise added, squared Euclidean model with noise interval $[0.2,0.4]$, RMSD model with noise interval $[0.2, 0.4]$, and random geometric model with noise interval $[0.3, 0.5]$. See panel \textit{(b)} for legend.}
    \label{fig:4}
\end{figure*}

We examine this \textit{overlapping} noise scenario by creating network models as before, but instead of switching from only model edges to only noise edges at $\rho_T$ (Fig.~\ref{fig:4}a, left), we now set an increasing noise interval $[\rho_a, \rho_b]$ (Fig.~\ref{fig:4}a, right). For edges added at densities $\rho< \rho_a$, model network edges are added in the usual ordering. If $\rho_a \leq \rho \leq \rho_b$, then with probability $p_{\rho} = \frac{1}{\rho_b - \rho_a}(\rho - \rho_a)$ we choose the next edge in the filtration at random and with probability $1-p_{\rho}$ we choose the next edge based on the ordered model edges. For $\rho>\rho_b$, all further edges are chosen at random.

We show all Betti curves generated by this process in Fig.~\ref{sfig:overlap} and highlight a few interesting results in Fig.~\ref{fig:4}b, c. First, because we have an expectation for edge density intervals with non-zero persistent homology, persistent cavities that are born exceptionally late can be considered significant \cite{sizemore2017cliques}. Following this concept, both the quantitative and qualitative evaluations of, for example the uniform configuration persistent homology output shown in Fig.~\ref{fig:4}b, left, could suggest interesting or significant features exist after $\rho=0.2$. Similar inferences might be drawn from the Betti curves of the geometric configuration model in Fig.~\ref{fig:4}b, right. However, we would be wrong to assign significance to the late born features in these plots. All of these late-born persistent cavities are generated by added noise.

Second, consider the Betti curve plots in Fig.~\ref{fig:4}c. Given the double peak of $\beta_1$ and $\beta_2$ \cite{sizemore2016classification}, one could reasonably conclude that all four have a considerable modular structure. However, again we would be wrong. Three of the four are distance-based models while only the assortative model (run without noise) actually contains distinct modules (Fig.~\ref{fig:4}c, left). The second peaks seen in the Betti curves generated by the distance-based models are solely produced by added noise, suggesting that qualitative interpretations of Betti curves arising from noisy data may be misleading.

Together, our results reveal the importance of a careful eye when interpreting the structure of weak edges in complex systems. Particularly, noise in the weak edges can falsely appear as significant structural features.

\section{Discussion}


In this work we investigated the structure of random edges added to pre-existing model network edges. We determined that the existing model structure dictates the topology of its added noise, and consequently that the structure of the added noise alone carries distinguishing information about the network model. We then identified generative processes for creating the three main patterns of added noise and finally highlighted consequences of variable noise structure for the analysis of weighted graphs.

\paragraph{Implications for data analysis}
The increasing use of weighted networks in applied sciences from molecular biology \cite{fuller2011review} to transportation \cite{xing2016weighted} suggests that weak edges and topological variability \cite{yan2019structural} will continue to be studied in the future. A major consequence of our results is that given the variable structure of added noise atop networks, extreme care must be taken when analyzing weighted networks with possible noise contamination. Though here we used persistent homology to query network structure, we expect that added noise structure as viewed by many graph metrics will vary based on real network topology, as seen previously in the binary case \cite{zalesky2016connectome,van2017proportional}. Consequently, we suggest that one considers the structure of noise dependent on their system as perceived by their structural measure of choice before assigning value to features formed by weak edges. On the other hand, our experiments suggest an avenue for improving the detection of a threshold (or range of densities) that separate data edges from noise edges \cite{wang2018network,zeng2012removing,zhou2012simplification}. If one knows the expected structure of added noise on their system, then one could use this information to determine at what threshold in a new dataset the noise begins.

\paragraph{Implications for theoretical work} Though motivated by problems in data analysis, our work also lays the foundation for additional interesting theoretical questions. First, one could interpret the experiments performed in this paper as querying the change of network structure caused by combining or shifting between two network models \cite{you2018graphrnn}. Here we combined one network model with random IID noise, but one could easily repeat these experiments with any pair of graph models. For example, how does the structure of a ring lattice combined with a modular network compare to that of a ring lattice combined with a random geometric model? Such questions could be helpful for understanding systems such as the brain that naturally switch between states \cite{bansal2019cognitive,cornblath2020temporal}. 

Second, an alternative way to interpret the patterns of added noise structure is that the model edges \emph{force} or \emph{restrict} the noise into a particular shape based on the topology of the empty space left by the arrangement of model edges. Following this line of thought, the \textit{random reversion} class of models could be seen as having a deferential structure, in that the added noise was quickly able to revert to its natural architecture, and the other two model classes could be interpreted as having a forceful structure, which dominates the ability of the added noise to revert. Indeed, we observed that even with only 15\% of edges added, the geometric configuration model has influenced the structure of the 85\% of randomly added edges, so that its added noise does not follow the IID noise pattern. We leave the question of how exactly the topology of the model edges dictate (or do not dictate) the topology of the added noise for future work. Additionally, one might ask if real-world sparse networks \cite{bassett2011conserved, naik2019sparse} have a dominant structure that protects their architecture against random fluctuations.

Third, given a sparse network, one can use randomly added edges and the crossover bar concept to help determine geometrical properties of the network's topological cavities \cite{obayashi2018volume}. Specifically, since we add edges at random, generating a distribution of death times for each topological cavity would suggest a cavity geometry that is more or less susceptible to randomly dying. One would expect cavities with large minimal generators to be unlikely to die via random edge addition, whereas cavities with multiple small minimal generators (for example a narrow tube) would likely die very soon from randomly adding edges since there are more opportunities to tessellate such a cavity.

\paragraph{Noisy networks and cognition} Finally, our experiments suggest interesting directions for future work in cognitive science and neuroscience. Studying the added noise structure is equivalent to studying the empty space left by a network, making the above analyses particularly interesting for systems in which the sparsity of edges is a feature. For example, in one's brain network the structure of both the present edges \emph{and} the empty space changes over development \cite{dennis2013development}. Specifically, network edges are pruned as a person ages and learns \cite{stephan2012complement}, which in turn alters the structure of the non-edges in their network \cite{morgan2018network}. Is the structure of the empty space in brain networks more or less important than the structure of those edges that exist? Finally, in cognitive science one could interpret learning as beginning with a noisy knowledge network in which some edges incorrectly connect disconnected concepts. Learning would then proceed by a rewiring that network to a final, correct knowledge network \cite{lynn2020abstract,melo2017detection}. Studying this phenomenon is effectively the reverse of the experiments presented here, in which one begins with a noisy network and ends with an expected model network.

\paragraph{Conclusion}
In conclusion, our work shows that the persistent homology of noise added to networks varies based on the real network topology. Additionally we find generative network rules that produce networks supporting differing structures of noise. Finally our results offer a reason to examine how the structure of added noise to a real network may influence the structural measure in question, in order to make real features present in weak edges clearly distinguishable from features created by noise alone.

\section{Methods}

 Computations were performed in julia, with the exception of the classification experiments which were performed in MATLAB. We use the Eirene software \cite{henselmanghrist16} for all persistent homology computations.

\subsection{Weighted network models}

We chose weighted network models that show a variety of real-world properties, including the structural feature of modularity, and the physical feature of weights that decreases with distance. For all models the final graph contained $N=70$ nodes, and 500 replicates were created. We chose these values for nodes and replicates to balance topological richness, reliable estimates of variance, and computation time. If a model yielded non-unique graph weights, random noise was added such that all edge weights would be unique and that the ranking of edges with unique edge weights would remain the same. See Section \ref{sec:net_generation} of the Supplementary Information and Fig.~\ref{sfig:graphinfo_1}, \ref{sfig:graphinfo_2}, \ref{sfig:graphinfo_3} for more details. We separate the descriptions of the models into two sections: distance-agnostic graph models and distance-based graph models.\\

\noindent \textbf{Distance-agnostic graph models.} The following models are created without any notion of a formal distance between nodes.

\begin{itemize}
    \item \textbf{Assortative}. The assortative model was constructed following an implementation of the weighted stochastic block model (WSBM) \cite{aicher2013adapting} in which four high-weight blocks were positioned along the diagonal.
    
    \item \textbf{Core periphery}. This model was also constructed with the WSBM \cite{aicher2013adapting} approach but high-weight blocks were positioned along the top and left edge of the adjacency matrix to form a core and periphery. One fourth of nodes formed the core, and the rest formed the periphery.
    
    \item \textbf{Disassortative}. The inverse of the assortative model, in which nodes within a community connect strongly to nodes outside of their community. 
    
    \item \textbf{Discrete uniform configuration}. A weighted configuration model with the node strengths drawn from the discrete uniform distribution.
    
    \item \textbf{Dot product}. Here we chose $N$ points at random in $\mathbb{R}^{dim}$ (here $dim=3$). We then weighted edges between two nodes as the dot product of the associated vectors.
    
    \item \textbf{Geometric configuration}. A weighted configuration model with node strengths drawn from a geometric distribution.
    
    \item \textbf{IID noise}. All edge weights were chosen at random from the uniform distribution on $(0,1)$.

\end{itemize}

\noindent \textbf{Distance-based graph models.} The following models are created by first choosing $N$ points in $\mathbb{R}^{dim}$, then calculating a distance $d(\Vec{u},\Vec{v})$ between every pair of points using the definitions below, and finally taking the reciprocal of that distance as the edge weight between two nodes. By this process, two nodes that are close together, as determined by the distance metric, will parent an edge with a large weight, whereas nodes that are far apart will parent edges with small weights. For simplicity, we chose $dim=3$ for all models.

\begin{itemize}

    \item \textbf{Cosine geometric}. Given nodes $v,u$ with associated vectors $\Vec{v}, \Vec{u} \in \mathbb{R}^{dim}$, respectively, the cosine distance is
    $$d(\Vec{v},\Vec{u}) =1 - \frac{ \Vec{v} \cdot \Vec{u}}{||\Vec{v}|| ||\Vec{u}||}.$$
    
    \item \textbf{Random geometric (Euclidean distance)}. Given nodes $v,u$ with associated vectors $\Vec{v}, \Vec{u} \in \mathbb{R}^{dim}$, respectively, $$d(\Vec{v},\Vec{u}) = \sqrt{\sum_{i=1}^{dim}(\Vec{v}_i - \Vec{u}_i)^2}.$$
    
     \item \textbf{Ring lattice}. We labeled $N$ vertices as $1,\dots, N$, and connected nodes in one large ring. We assigned the edge weight between nodes $i$ and $j$ as the inverse hop distance along this ring, and assumed that hopping is allowed only between neighboring nodes \cite{watts1998collective}. 
    
    \item \textbf{Root mean squared deviation (RMSD)}. Given nodes $v,u$ with associated vectors $\Vec{v}, \Vec{u} \in \mathbb{R}^{dim}$, respectively, $$d(\Vec{v},\Vec{u}) = \sqrt{\frac{1}{dim} \sum_{i=1}^{dim}(\Vec{v}_i - \Vec{u}_i)^2}.$$

    \item \textbf{Squared Euclidean distance}. Given nodes $v,u$ with associated vectors $\Vec{v}, \Vec{u} \in \mathbb{R}^{dim}$, respectively, $$d(\Vec{v},\Vec{u}) = \sum_{i=1}^{dim}(\Vec{v}_i - \Vec{u}_i)^2.$$

\end{itemize}

\subsubsection{Models to reproduce noise profiles}

We created three models with the intention of finding simple rules that would give rise to a particular added noise persistent homology pattern. First, the weighted probabilistic triangle graph takes one parameter $p$ that controls triangle formation. The goal of this model is to form a weighted network such that when we expand to the filtration, at each step in the filtration the new edge has probability $p$ of forming a triangle. Beginning with $N$ nodes and 0 edges, with probability $p$ we either add an edge that will create a new triangle, or we add an edge at random. If we are to add a new triangle, we check to see if there are any open triangles in the graph -- that is where two edges connect three nodes -- and if there are no open triangles, we add an edge at random. If there are open triangles, we pick one open triangle to fill with probability proportional to the product of the two edge weights of the open triangle edges. The new edge is assigned a weight lower than any previously added edge.

The second and third models are complementary to the first. In the second model, the random $m$-clique networks take one parameter $m$ that controls the size of cliques to be added. Beginning with an empty network we randomly choose $m$ nodes and add a value of 1 to each edge weight connecting the $m$ nodes. This process repeats until no more than 12\% of edges were empty. In the third model, we create a weighted clique model in which throughout the filtration, each new edge contributes to forming one growing clique. For example, once three edges have been added the graph is a $3$-clique, the first 6 edges will form a $4$-clique, the first 10 edges added will form a $5$-clique, and so on.

\subsection{Persistent homology} 

Persistent homology \cite{edelsbrunner2008persistent,otter2017roadmap} measures the birth and death of persistent cavities that arise and evolve throughout a sequence of simplicial complexes in which simplices may be added at each step (a filtered simplicial complex). Here we form this sequence from a weighted network by creating a graph filtration 
\begin{equation} 
G_0 \subseteq G_1 \subseteq \dots  \subseteq G_{|E|},
\end{equation}
where $G_i$ is the binary graph containing edges with the $i$ highest weights in the weighted network, and then taking the clique complex of each $G_i$ \cite{giusti2015clique,petri2013topological,horak2009persistent,kahle2009topology}. We compute the persistent homology using the Eirene \cite{henselmanghrist16} package in julia. See the Supplementary Methods in the Appendix for more details.

\subsubsection{Derived values}

Following Ref. \cite{adcock2012ring}, we use the three barcode summaries listed below. Intuitively, each returns a description of the amount of persistent homology in each dimension, but they vary by their weighting of each bar $[b_l, d_l)$ of the barcode. 

\begin{itemize}
    \item \textbf{Betti bar, $\bar{\beta}_k$}. Let $M$ be the total number of persistent cavities in dimension $k$. Then 
    $$\bar{\beta}_k = \sum_{l=1}^M(d_l - b_l).$$ The $\bar{\beta}_k$ value sums the lifetimes of all bars in dimension $k$.
    
    \item \textbf{Mu bar, $\bar{\mu}_k$}. Let $M$ be the total number of persistent cavities in dimension $k$. Then 
    $$\bar{\mu}_k = \sum_{l=1}^Mb_l(d_l - b_l).$$ The $\bar{\mu}_k$ value scales each bar's lifetime by the birth time and then sums these weighted lifetimes.
    
    \item \textbf{Nu bar, $\bar{\nu}_k$}. Let $M$ be the total number of persistent cavities in dimension $k$ and $L$ the number of edges in the complete graph. Then 
    $$\bar{\nu}_k = \sum_{l=1}^M(L-d_l)(d_l - b_l).$$ The $\bar{\nu}_k$ scales each bar's lifetime based on the death time of that bar, and then sums the scaled lifetimes.
    
\end{itemize}

\subsection{Classification}

We seek a simple and generative method to classify our networks to ensure flexibility in incorporating new network models or data that may be generated from a different underlying distribution. As such, we model the distribution of features for each network model using a multivariate Gaussian model, and collect these models into a Gaussian mixture model \cite{reynolds2009gaussian}. For prediction, we use features from a held-out test set of network features, and assign a predicted label based on the class that generates the highest posterior probability. These predicted labels are then used to generate the confusion matrices of the main text.

\subsubsection{Generative Gaussian Mixture Model}

For each network instantiation $i$ belonging to network model $j$, we collect a 12-dimensional vector $\bm{x}^i_j$ of features $\bar{\beta}_k$, $\bar{\mu}_k$, and $\bar{\nu}_k$ for $k = 1,2,3,4$. Then, we compute the mean $\bm{\mu}_j$ and covariance $\Sigma_j$ of the features for the $N_\mathrm{train} = 250$ networks belonging to model $j$, to yield the Gaussian probability distribution of network model $j$ as
$$ G_j(\bm{x}|\mu_j,\Sigma_j) = \frac{1}{\sqrt{2\pi |\Sigma_j|}}e^{-\frac{1}{2}(\bm{x} - \bm{\mu}_j)^\top \Sigma_j^{-1} (\bm{x} - \bm{\mu}_j)}. $$
Because some features are all 0 at some thresholds, we bias the covariance matrix by adding $0.01I$ to ensure $\Sigma_j$ is positive definite. Next, we collect these models together into the following equally weighted Gaussian mixture:
$$ p(\bm{x}) = \sum_{j=1}^J \frac{1}{J} G_j(\bm{x}|\bm{\mu}_j,\Sigma_j), $$
where $J = 12$ is the total number of network models we used. Finally, for a test set of $N_{\mathrm{test}} = 250$ for each model class $j$ (for a total of $JN_{\mathrm{test}} = 3000$ network instantiations), we computed the posterior probability for each of the mixture components, and assigned a label to each network corresponding to the class of its maximum posterior probability. While the confusion matrices represent one selection of training and testing sets, we measure the distribution of performance by taking the total classification accuracy for each feature set across $100$ random and disjoint sets of training and testing sets (see Fig.~\ref{sfig:classification_accuracy}).

\subsubsection{Features}

Below we detail the inputs to each of the eight classification experiments performed in this work. For each, we take a specific subset of barcodes and use them to compute $\bar{\beta}_k$, $\bar{\mu}_k$, and $\bar{\nu}_k$. Recall that the barcode in dimension $k$ is a set of pairs $(b_l,d_l)$ corresponding to the birth and death time of persistent cavity $i$, respectively.

\begin{itemize}
    \item \textbf{Added noise (in context).} We computed persistent homology on the entire graph sequence from $\rho=0$ to $\rho=1$, where at $\rho = \rho_T$ edge additions became random. To understand the information contained in the added noise portion of the barcode, we replaced any $b_l, d_l < \rho_T$ with $\frac{\rho_T|E|+1}{|E|}$ (the density at which the first noisy edge is added). Any cavity with $b_l = d_l$ will contribute 0 to each of the three summary statistics. Any bar with $b_l<\rho_T$ and $d_l>\rho_T$ will effectively be shortened so that the persistent cavity is born at $\frac{\rho_T|E|+1}{|E|}$.
    
    \item \textbf{Model weights before $\rho_T$.} From the barcodes computed using the full filtration, we replaced any $b_l, d_l > \rho_T$ with $\rho_T$. Any bar with $b_l<\rho_T$ and $d_l > \rho_T$ will effectively be shortened so that it dies at $\rho_T$. 
    
    \item \textbf{All barcodes.} Here we kept the barcodes as they were originally calculated.
    
    \item \textbf{Added noise (isolated).} Persistent homology was calculated on the added noise graph alone so we used these barcodes for the classification.
    
    \item \textbf{Crossover bars.} We filtered the added noise barcode to only include bars with original birth $b_l < \rho_T$ and original death $d_l > \rho_T$. Because we begin with the added noise portion of the barcode, the smallest $b_l$ possible is $\frac{\rho_T|E|+1}{|E|}$.
    
    \item \textbf{Noise exclusive bars.} We filtered the barcode to only include bars with $b_l >\rho_T$ and $d_l >\rho_T$.
    
    \item \textbf{Randomized model edge weights.} First, we randomized the ordering of the model edges by randomizing their weights. We then computed the persistent homology on this randomized model, and finally we replaced every $b_l, d_l > \rho_T$ with $\rho_T$, so that any bar with $b_l <\rho_T$ and $d_l>\rho_T$ was effectively shortened so that it would die at $\rho_T$.

\end{itemize}

\section{Data Availability}
All data can be generated using the open code hosted at \url{https://github.com/asizemore/Noise_and_TDA}.

\section{Code Availability}
All code to generate data and perform analyses can be found at \url{https://github.com/asizemore/Noise_and_TDA}. Interactive visualizations are hosted at \url{https://asizemore.github.io/noise_and_tda_supplement/}.

\section{Acknowledgments}
The authors especially thank Dr. Erin Teich, Dr. Linden Parkes, Darrick Lee, Dr. Jakob Hansen, Zoe Cooperband, and Dr. Lia Papadopolous for their helpful comments and insightful feedback. This work was funded by the Army Research Office through contract number W911NF-16-1-0474. DSB and ASB also acknowledge additional support from the John D. and Catherine T. MacArthur Foundation, the Alfred P. Sloan Foundation, the
ISI Foundation, the Paul Allen Foundation, the Army Research Laboratory (W911NF-10-2-0022), the Army Research Office (Bassett-W911NF-14-1-0679, DCIST- W911NF-17-2-0181), and the National Science Foundation (NSF PHY-1554488, BCS-1631550, and IIS-1926757). The content is solely the responsibility of the authors and does not necessarily represent the official views of any of the funding agencies.

\section{Citation Diversity Statement}
Recent work in several fields of science has identified a bias in citation practices such that papers from women and other minority scholars are under-cited relative to the number of such papers in the field \cite{mitchell2013gendered,maliniak2013gender,caplar2017quantitative,dion2018gendered,dworkin2020extent}. Here we sought to proactively consider choosing references that reflect the diversity of the field in thought, form of contribution, gender, race, ethnicity, and other factors. First, we obtained the predicted gender of the first and last author of each reference by using databases that store the probability of a first name being carried by a woman \cite{dworkin2020extent,zhou2020gender}. By this measure (and excluding self-citations to the first and last authors of our current paper), our references contain 14.79\% woman(first)/woman(last), 6.37\% man/woman, 23.63\% woman/man, and 55.22\% man/man. This method is limited in that a) names, pronouns, and social media profiles used to construct the databases may not, in every case, be indicative of gender identity and b) it cannot account for intersex, non-binary, or transgender people.

\onecolumn


\bibliographystyle{plain}
\bibliography{bibfile.bib}

\newpage
\setcounter{figure}{0}
\renewcommand{\thefigure}{A\arabic{figure}}
\section{Supplementary Information}

\subsection{Supplementary Methods}

Persistent homology measures higher order structure in weighted networks by detecting topological cavities that form at different edge weight (or density) thresholds. Those cavities that exist across thresholds are called persistent cavities. The three main steps of persistent homology for weighted graphs as performed in this paper are (i) creating a sequence of binary graphs from the weighted network, (ii) transforming this sequence of binary graphs into a sequence of simplicial complexes, and (iii) detecting persistent cavities that form and collapse throughout the sequence \cite{giusti2015clique,petri2013topological,horak2009persistent,kahle2009topology}. We discuss these three steps in more detail below, but we advise the interested reader to also consult Refs. \cite{ghrist2008barcodes,otter2017roadmap,petri2013topological}. 

\subsubsection{From weighted network to sequence of binary graphs} Given a weighted network in which all edge weights are unique, the edge weights induce an ordering on the edges from greatest to least. We can follow this ordering to create a sequence of binary graphs with one new edge added at each step. More rigorously, we construct 

\begin{equation} \label{eq:graph_filtration}
G_0 \subseteq G_1 \subseteq \dots  \subseteq G_{|E|} ,
\end{equation}

\noindent where $G_i$ is the binary graph containing the strongest (i.e. highest weighted) $i$ edges in the original weighted network. In our case, we always have $|E| = \binom{N}{2}$ so that the last graph $G_{|E|}$ is the complete graph on $N$ nodes.

\subsubsection{From sequence of binary graphs to sequence of simplicial complexes} Next we transform our graphs into simplicial complexes so that we can detect higher order structure. A simplicial complex is similar to a graph in that it records connectivity between nodes, but in a simplicial complex a set of $k$ nodes can participate in a polyadic relation called a \emph{simplex}. Specifically, a simplicial complex is a set $\mathcal{K} = (V,S)$ where $V$ is the vertex set and $S$ is the set of simplices, such that if $\sigma \subset V$ is a simplex ($\sigma \in S$), then for any $\tau \subseteq \sigma$, $\tau$ is also a simplex ($\tau \in S$). Geometrically, a $k$-simplex is the convex hull of $k+1$ affinely positioned points, which we interpret as a building block on $k+1$ nodes within the complex. Intuitively, a 0-simplex is a node, a 1-simplex an edge, a 2-simplex a filled triangle, a 3-simplex a filled tetrahedron, and so on.

We can transform a binary graph $G$ into a simplicial complex $X(G)$ by adding a $k$-simplex between $k+1$ nodes whenever the $k+1$ nodes are all-to-all connected (form a $k+1$ clique). The constructed $X(G)$ is called the \emph{clique complex} or \emph{flag complex} of $G$. Now that we can take an arbitrary binary graph $G$ and create from it a simplicial complex $X(G)$, we perform this step on all binary graphs in Eq. \ref{eq:graph_filtration}. The result is a sequence of simplicial complexes 

\begin{equation} \label{eq:sc_filtration}
    X(G_0) \subseteq X(G_1)) \subseteq \dots  \subseteq X(G_{|E|}) ,
\end{equation}

\noindent where each $X(G_i)$ is the clique complex of $G_i$. 

\subsubsection{Extracting persistent cavities} Beginning with one simplicial complex $X(G_i)$, a topological cavity of dimension $k>0$ is a shell of $k$-simplices that is not filled by $(k+1)$-simplices. A cavity in dimension 1 could manifest as a ring of 1-simplices (edges), the interior of a loop of 2-simplices such that the inside is empty, or even a tube of 2-simplices. In dimension 2, a topological cavity could similarly manifest as a shell of 2-simplices (such as an octagon), or perhaps as a shell of 2-simplices with many 3-simplices attached on the outside. Cavities in dimensions $>2$ are higher dimensional analogs of the given examples.

Persistent homology does not simply count the topological cavities in each dimension, it also \emph{tracks} the cavities throughout the sequence of simplicial complexes. Consider the step $X(G_i) \rightarrow X(G_{i+1})$ where the $\rightarrow$ indicates the mapping sending nodes and simplices in $X(G_i)$ to their counterparts in $X(G_{i+1})$. With the addition of the $i+1$ edge, three non-exclusive topological situations could occur. First, the edge addition could create a new $k$-simplex that completes the shell of a $k$-cavity. In other words, this new edge could form a $k$-cavity. Second, some $k$-cavity that existed in $X(G_i)$ may still be a cavity (non-tessellated shell), although it might be smaller given the addition of simplices added with the new edge. In this case we say that the $k$-cavity persists from $X(G_i)$ to $X(G_{i+1})$. Third and finally, the new edge may add simplices such that a cavity that existed in $X(G_i)$ becomes tessellated, or filled in, in $X(G_{i+1})$. Here we say that edge $i+1$ killed the $k$-cavity. When we extend this process across the entire sequence of simplicial complexes from Eq. \ref{eq:sc_filtration}, we recover the formation, persistence, and tessellation of topological cavities. The collection of persistent topological cavities is called the persistent homology of the weighted network.

\subsubsection{Barcodes and Betti curves} For any persistent cavity, the edge density at which the cavity first appears is called the \emph{birth}, and the edge density at which the cavity is killed is called the \emph{death}. The persistent homology records the (birth, death) pair of every persistent cavity. The collection of persistent $k$-cavities is called the \emph{barcode} and is visually represented as a sequence of horizontal lines in which each line represents one persistent cavity and extends from the persistent cavity birth to its death.  

Barcodes as mathematical objects lack particular helpful properties, such as a unique mean, so we often represent the persistent homology in dimension $k$ as a Betti curve $\beta_k$. At edge density $\rho$, $\beta_k(\rho)$ is the number of persistent cavities alive at density $\rho$. Though Betti curves do not retain the persistence information of the barcodes, they are helpful for visualization and interpretation of the persistent homology of graph models because we can compute the mean and standard error across replicates.

\subsubsection{Model network generation details}\label{sec:net_generation}

To supplement our descriptions in the Methods section, we here provide extended details regarding our generation of network models. As stated above, every network contained $N=70$ nodes and from each model network we generated 500 replicates. If the edge weights were not unique, random noise was added to ensure uniqueness while retaining the relative ordering of edges.\\

\textit{\textbf{IID noise}} Each edge weight was drawn uniformly at random from $[0, 1]$.\\

\textit{\textbf{Weighted stochastic block model networks}} Using code from \cite{aicher2013adapting} (rewritten in julia), we created the assortative, core periphery, and disassortative models. All three models consisted of different arrangements of 16 blocks describing edges between $N/4$ nodes (or approximately $N/4$ nodes if the integer $N$ is not divisible by 4). Each block was either a high-weight or a low-weight block. The high-weight blocks consisted of entries drawn from a normal distribution with parameters $\mu=20$, $\sigma=5$ for the assortative and disassortative models, and $\mu=15$, $\sigma=5$ for the core periphery model. The low-weight blocks contained entries drawn from a normal distribution with $\mu=10$ and $\sigma=5$.

High- and low-weight blocks were arranged based on the specific model type. For the assortative model, four high-weight blocks were placed along the diagonal, and the rest of the matrix was filled with low-weight blocks. For the disassortative model, four low-weight blocks were placed on the diagonal and the rest of the matrix was filled with high-weight blocks. Finally, the core periphery model contained high-weight blocks along the top row (of four rows) and along the first column (of four columns) of blocks, while the rest was filled with low-weight blocks. Additionally, see Fig.~\ref{sfig:graphinfo_1} for block arrangements.\\

\textit{\textbf{Configuration models}} For both the geometric configuration and discrete uniform configuration models, we formed networks using julia code inspired by Ref. \cite{newman2003structure} and the NetworkX implementation \cite{hagberg2008exploring}. To create a network, we first drew node strengths from a distribution and set this vector as the target node strength vector; for the geometric configuration model, the chosen distribution was a geometric distribution with $p=0.01$ then scaled by 100, whereas for the discrete uniform configuration model, the chosen distribution was a discrete uniform distribution with parameters $a=0$, $b=1000$. Beginning with an empty graph, we repeatedly joined pairs of randomly chosen nodes with edges of weight 1, counting $l$ connections formed between a pair of nodes as one edge with weight $l$. This process continued until the strength of each node in the network matched the target node strength defined by the target node strength vector. See Refs. \cite{newman2003structure,hagberg2008exploring} for more details and \url{https://github.com/asizemore/Noise_and_TDA} for code.\\

\textit{\textbf{Dot product}} To create the dot product network, we chose $N$ points uniformly at random from $\mathbb{R}^{dim}$. Here, $N=70$ and $dim=3$. Each point $\Vec{v}$ chosen in $\mathbb{R}^{dim}$ was associated with a node $v$ in the network. We then assigned the edge weight between $v$ and $u$ to be $\Vec{v} \cdot \Vec{u}$.\\

\textit{\textbf{Distance-based models}} The five distance based models rely on different distance metrics to determine edge weights. Below we detail each in turn. For all except the weighted ring lattice, the model begins with choosing $N$ points $\Vec{v}$ uniformly at random from $\mathbb{R}^{dim}$, and associating these points ($\Vec{v}$) with nodes ($v$). For the results reported in this paper, $dim=3$. Each model has an associated distance metric $d(\Vec{v},\Vec{u})$, and the edge weight between nodes $v, u$ for the model are assigned to $\frac{1}{d(\Vec{v},\Vec{u})}$. See below for the chosen distance functions.

\begin{itemize}
    \item For the cosine geometric model,
    $$d(\Vec{v_i},\Vec{v_j}) =1 - \frac{ \Vec{v_i} \cdot \Vec{v_j}}{||\Vec{v_i}|| ||\Vec{v_j}||}.$$
    
    \item For the random geometric model, $$d(\Vec{v},\Vec{u}) = \sqrt{\sum_{i=1}^{dim}(\Vec{v}_i - \Vec{u}_i)^2}.$$
    
    \item For the root mean squared deviation model, $$d(\Vec{v},\Vec{u}) = \sqrt{\frac{1}{dim} \sum_{i=1}^{dim}(\Vec{v}_i - \Vec{u}_i)^2}.$$
    
    \item For the squared euclidean distance model, $$d(\Vec{v},\Vec{u}) = \sum_{i=1}^{dim}(\Vec{v}_i - \Vec{u}_i)^2.$$

\end{itemize}

Finally, forming the weighted ring lattice network can be imagined as placing $N$ nodes uniformly around a circle, and then connecting node pairs with edges weighted according to their distance. Specifically, given nodes $v_1, \dots, v_N$, we connect all pairs $(v_i,v_j)$, $i<j$ by an edge with weight 1 if $j-i=1$ or $i=1$, $j=N$. The distance $d(v_i,v_j)$ is then the hop distance between nodes $v_i$ and $v_j$ on this ring. For the final weighted network, we assign edge weights to equal $\frac{1}{d(v_i,v_j)}$ between nodes $v_i$ and $v_j$. For example, the edge connecting $v_3$ and $v_{N=70}$ in our networks has a weight of $1/3$ \cite{watts1998collective}.

\subsection{Supplementary Results}

Below we include additional plots and results that are not shown in the main text, but that serve to support our observations and inferences.

\subsubsection{Properties of model networks}

In order to investigate added noise from a variety of perspectives, we chose network models with a wide range of properties. In Fig.~\ref{sfig:graphinfo_1}, \ref{sfig:graphinfo_2}, \ref{sfig:graphinfo_3} we show an example weighted network created from each model (top). We additionally show the strength distribution (middle) and edge weight distribution (bottom) of one weighted network serving as an example of the given model.

\begin{figure}
    \centering
    \includegraphics[width=\linewidth]{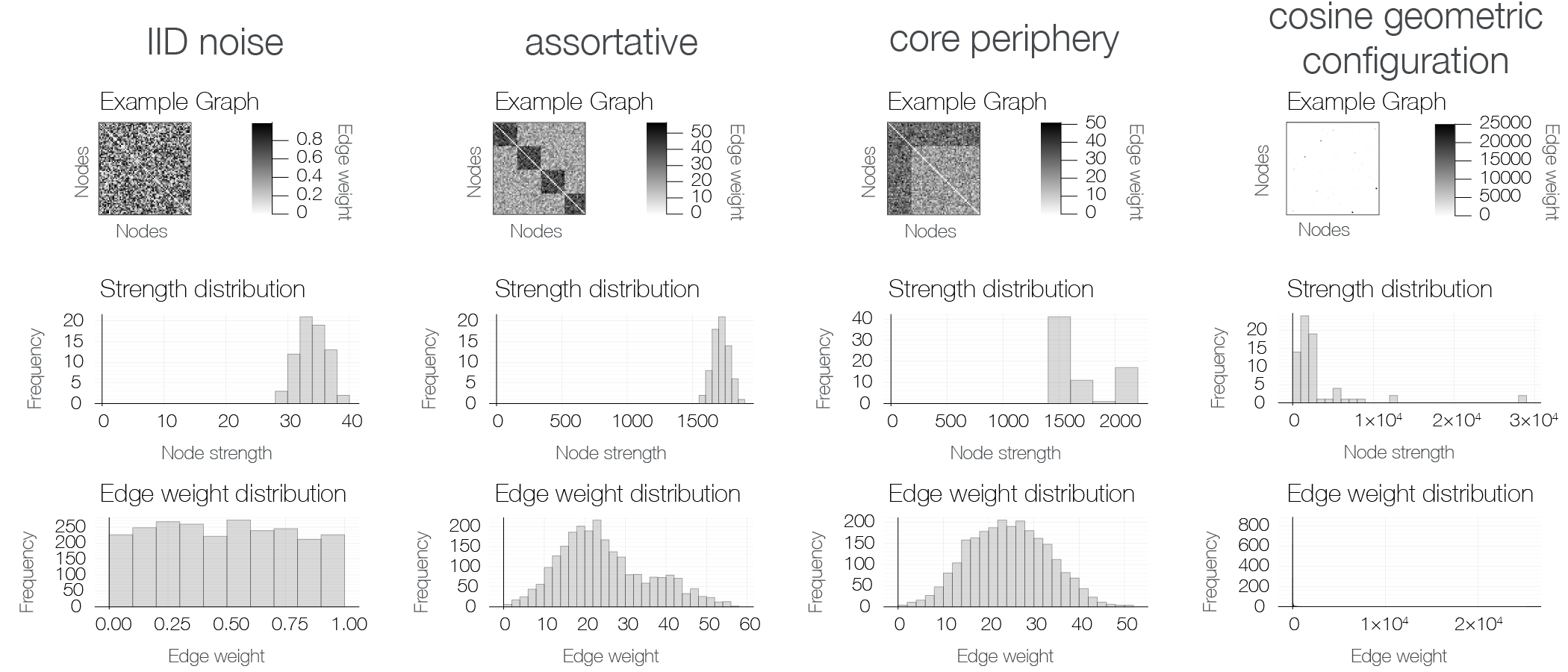}
    \caption{\textbf{Network properties for the first four of twelve models.} For each model, we show the adjacency matrix (top), node strength distribution (middle), and edge weight distribution (bottom) of an example graph.}
    \label{sfig:graphinfo_1}
\end{figure}

\begin{figure}
    \centering
    \includegraphics[width=\linewidth]{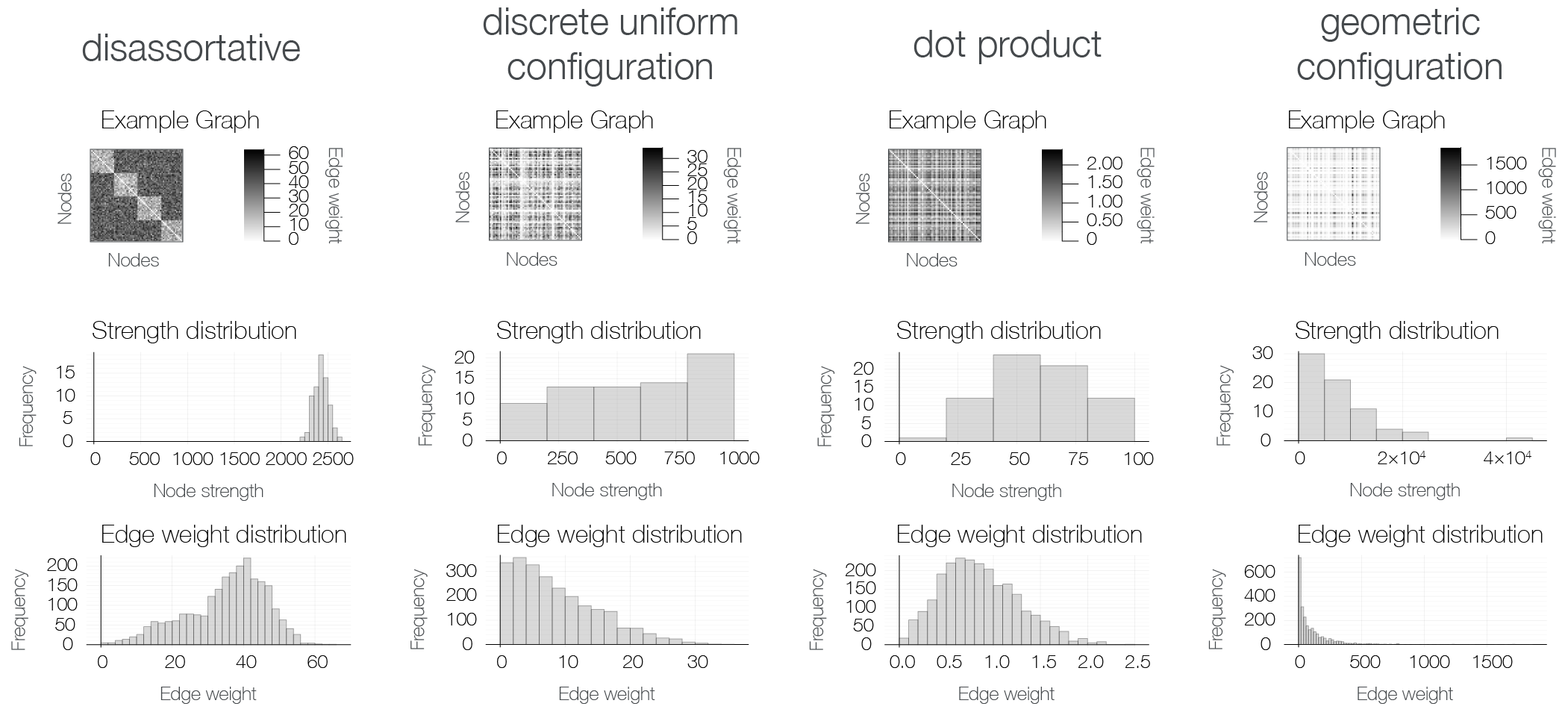}
    \caption{\textbf{(Continued from previous) Network properties for the middle four of twelve models.} For each model, we show the adjacency matrix (top), node strength distribution (middle), and edge weight distribution (bottom) of an example graph.}
    \label{sfig:graphinfo_2}
\end{figure}

\begin{figure}
    \centering
    \includegraphics[width=\linewidth]{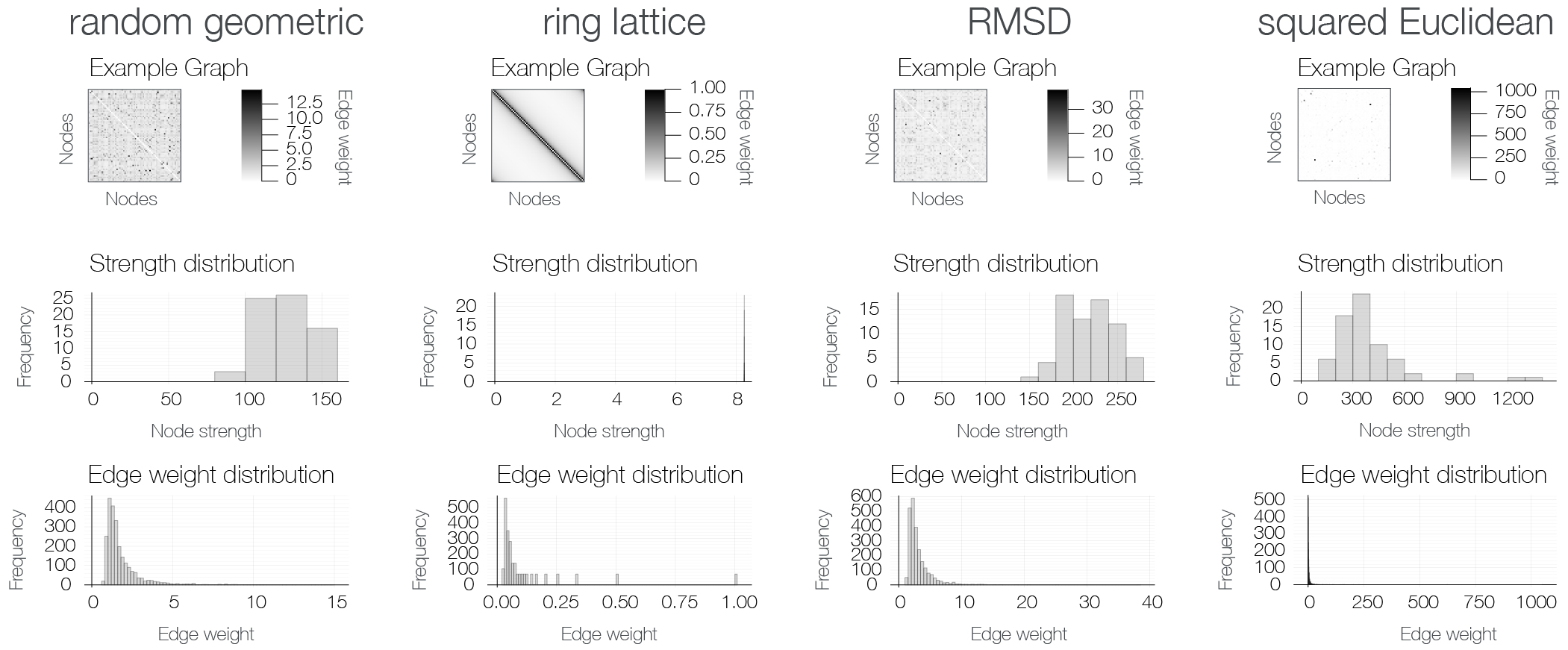}
    \caption{\textbf{(Continued from previous) Network properties for the final four of twelve models.} For each model, we show the adjacency matrix (top), node strength distribution (middle), and edge weight distribution (bottom) of an example graph.}
    \label{sfig:graphinfo_3}
\end{figure}

\clearpage
\newpage
\subsubsection{Additional Betti curve plots} \label{sec:betti_curves_supp}

\begin{figure}
    \centering
    \includegraphics{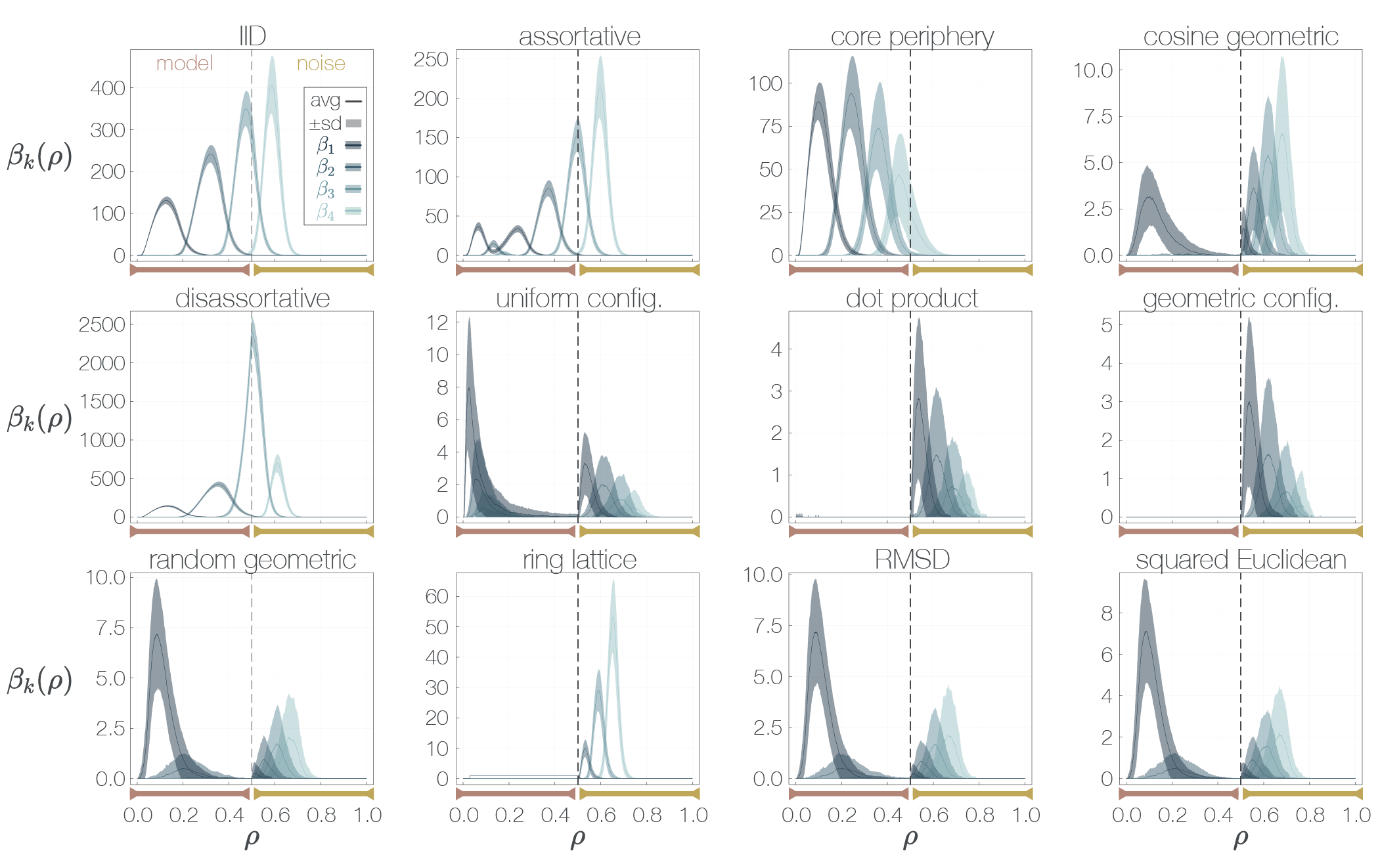}
    \caption{\textbf{Betti curves for all models at $\rho_T=0.5$} Average Betti curves in dimensions 1-4 shown in solid lines and $\pm$ the standard deviation shown with shaded regions. See Fig.~\ref{fig:1} for more details.}
    \label{sfig:bettis_05}
\end{figure}

To supplement Fig.~\ref{fig:1}, we include the Betti curves for all models with $\rho_T = 0.5$ in Fig.~\ref{sfig:bettis_05}. As in previous figures, we include Betti curves in dimensions 1-4, averaged over 500 replicates, as well as the standard deviation. The rose portion of the x-axis indicates where the filtration follows the model weights ordering, while the gold portion of the x-axis indicates the portion of the filtration that adds edges randomly.

Below we show Betti curves for all of the tested values of $\rho_T$ for all experiments discussed in the main text. Given the large number of replicates, experimental setups, and values of $\rho_T$ tested, we additionally host interactive visualizations at \url{https://asizemore.github.io/noise_and_tda_supplement/}. 

\begin{figure}[hbtp!]
    \centering
    \includegraphics[width=0.9\textwidth]{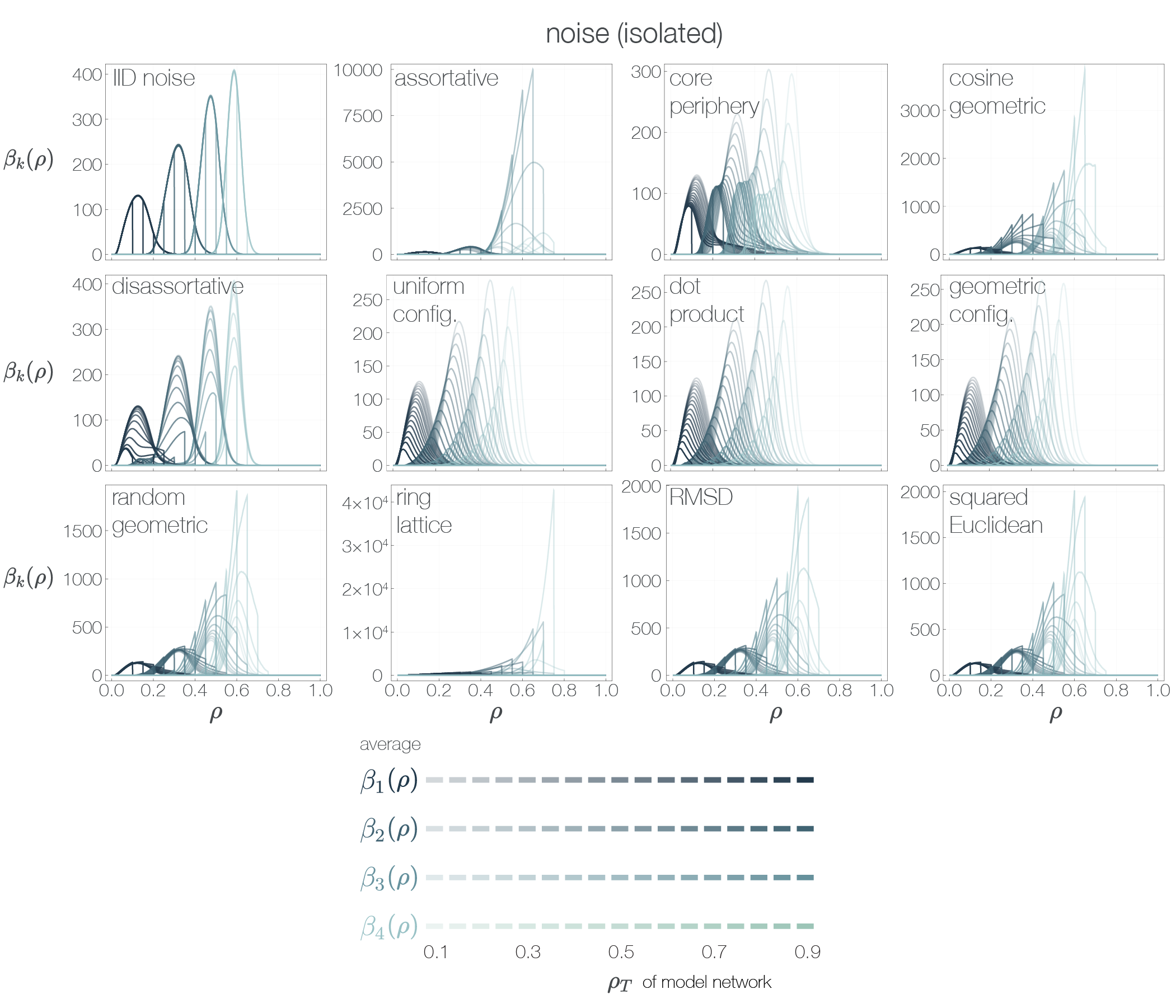}
    \caption{\textbf{Betti curves generated from isolated noise networks across values of $\rho_T$.} See Fig.~\ref{fig:1} for plot details.}
    \label{sfig:noiseonly}
\end{figure}

In Section \ref{sec:classification} we took the added noise portion of the thresholded model network runs and computed the persistent homology of the added noise network in isolation. Said another way, after constructing the gold section of Fig.~\ref{fig:intro}, we took all of the gold edges to be one weighted network. If the model was thresholded at $\rho_T$, the isolated noise network would have density $1-\rho_T$. In the same style as Fig.~\ref{fig:1}c, we show the average isolated noise network Betti curves for all tested values of $\rho_T$ in Fig.~\ref{sfig:noiseonly}.

\begin{figure}[hbtp!]
    \centering
    \includegraphics[width=0.9\textwidth]{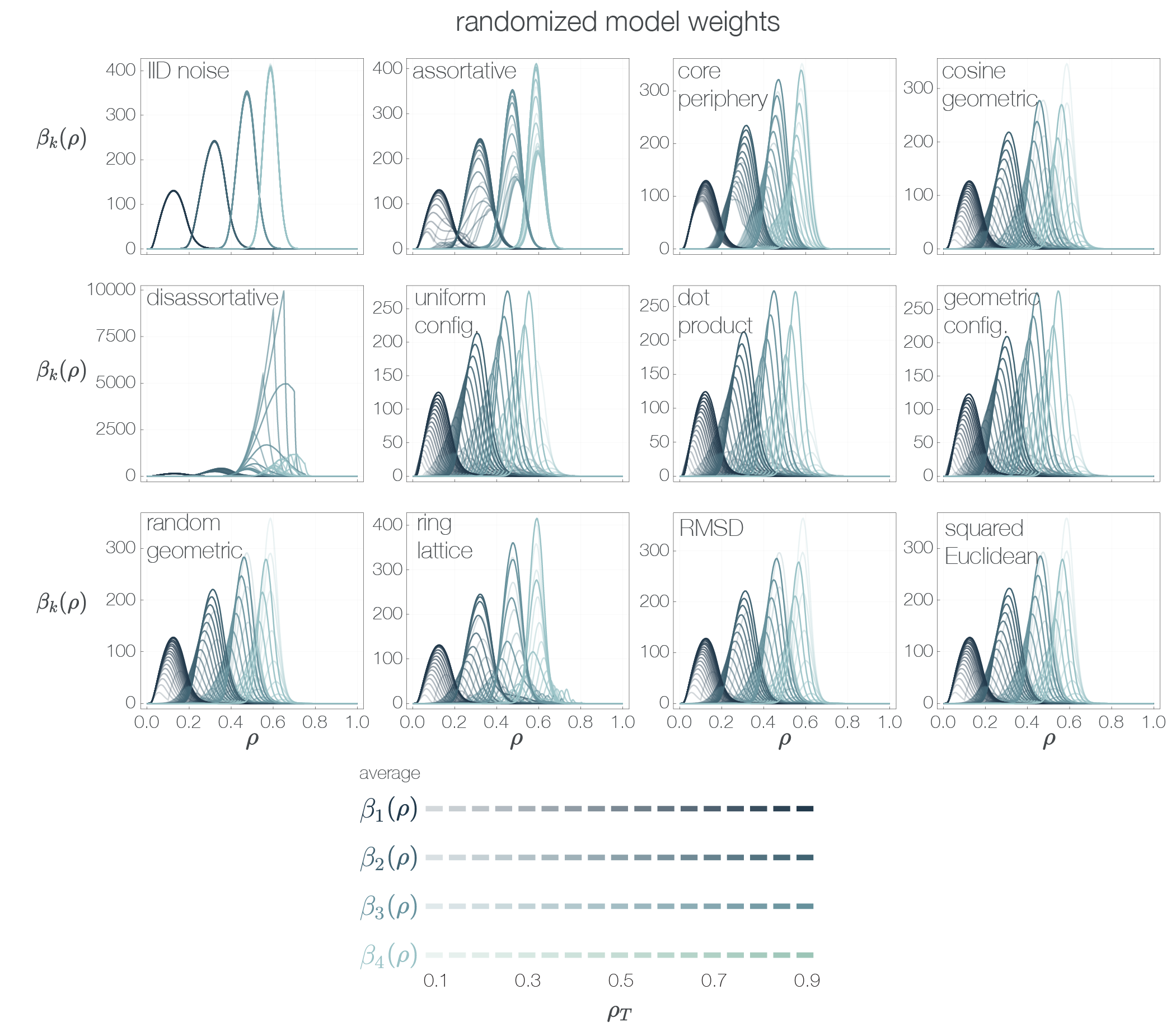}
    \caption{\textbf{Betti curves generated from randomized model weight experiments across values of $\rho_T$.} See Fig.~\ref{fig:1} for plot details.}
    \label{sfig:randomized}
\end{figure}

\clearpage
\newpage
In Section \ref{sec:classification} we also investigated how randomizing the model edge weights would affect the classification results. In Fig.~\ref{sfig:randomized} we show the average Betti curves generated by randomizing model weights across all tested values of $\rho_T$.

\begin{figure}[hbtp!]
    \centering
    \includegraphics[width=0.9\textwidth]{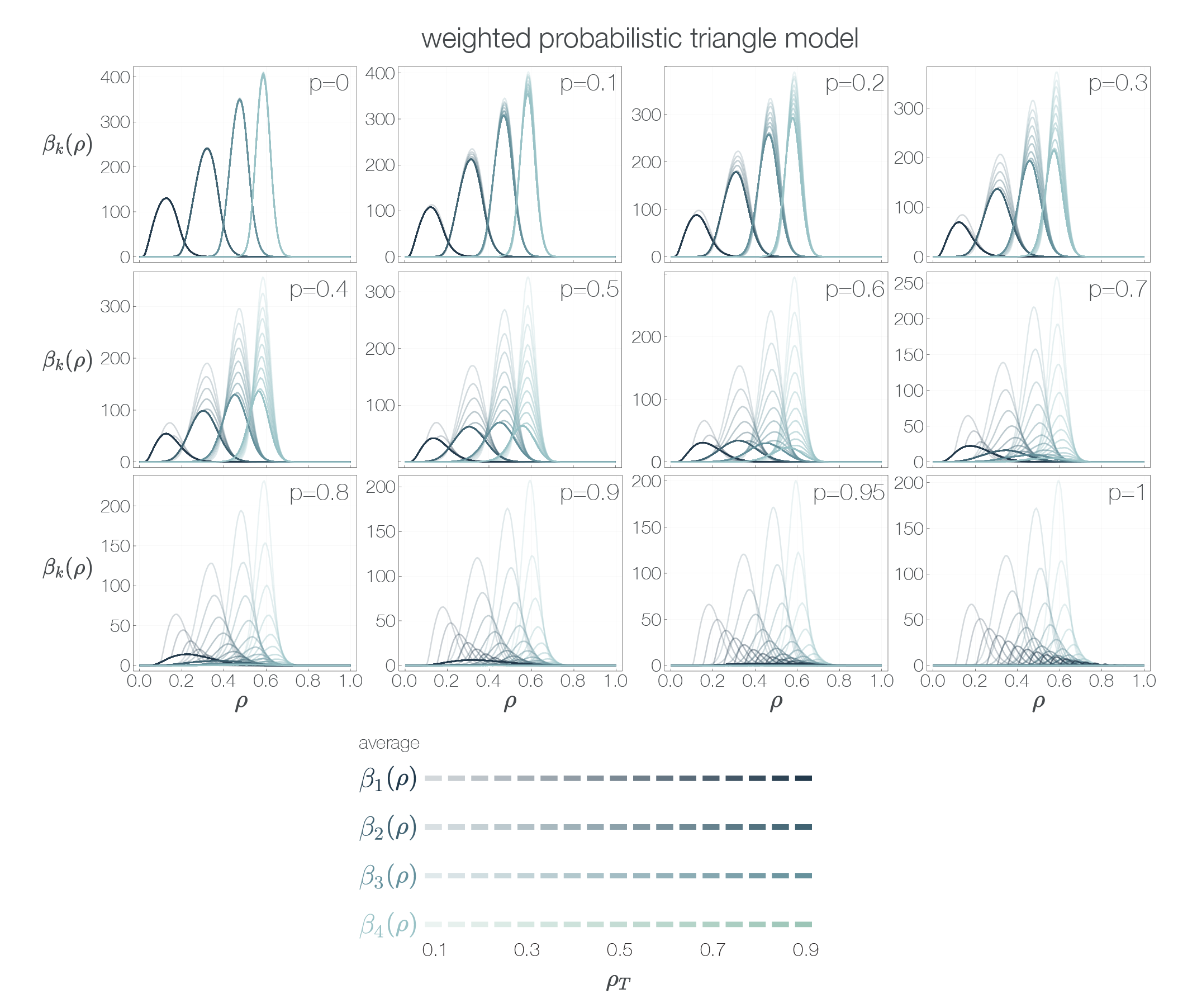}
    \caption{\textbf{Betti curves generated from the weighted probabilistic triangle model across values of $p$, $\rho_T$.} See Fig.~\ref{fig:1} for plot details.}
    \label{sfig:triangle}
\end{figure}

\clearpage
\newpage
As discussed in Section \ref{sec:drivers}, the weighted probabilistic triangle model produces a network in which the value of the parameter $p$ affects the likelihood of forming triangles as the filtration unfolds. We show average Betti curves across tested values of $\rho_T$ for all investigated values of $p$ in Fig.~\ref{sfig:triangle}.

\begin{figure}[hbtp!]
    \centering
    \includegraphics[width=0.9\textwidth]{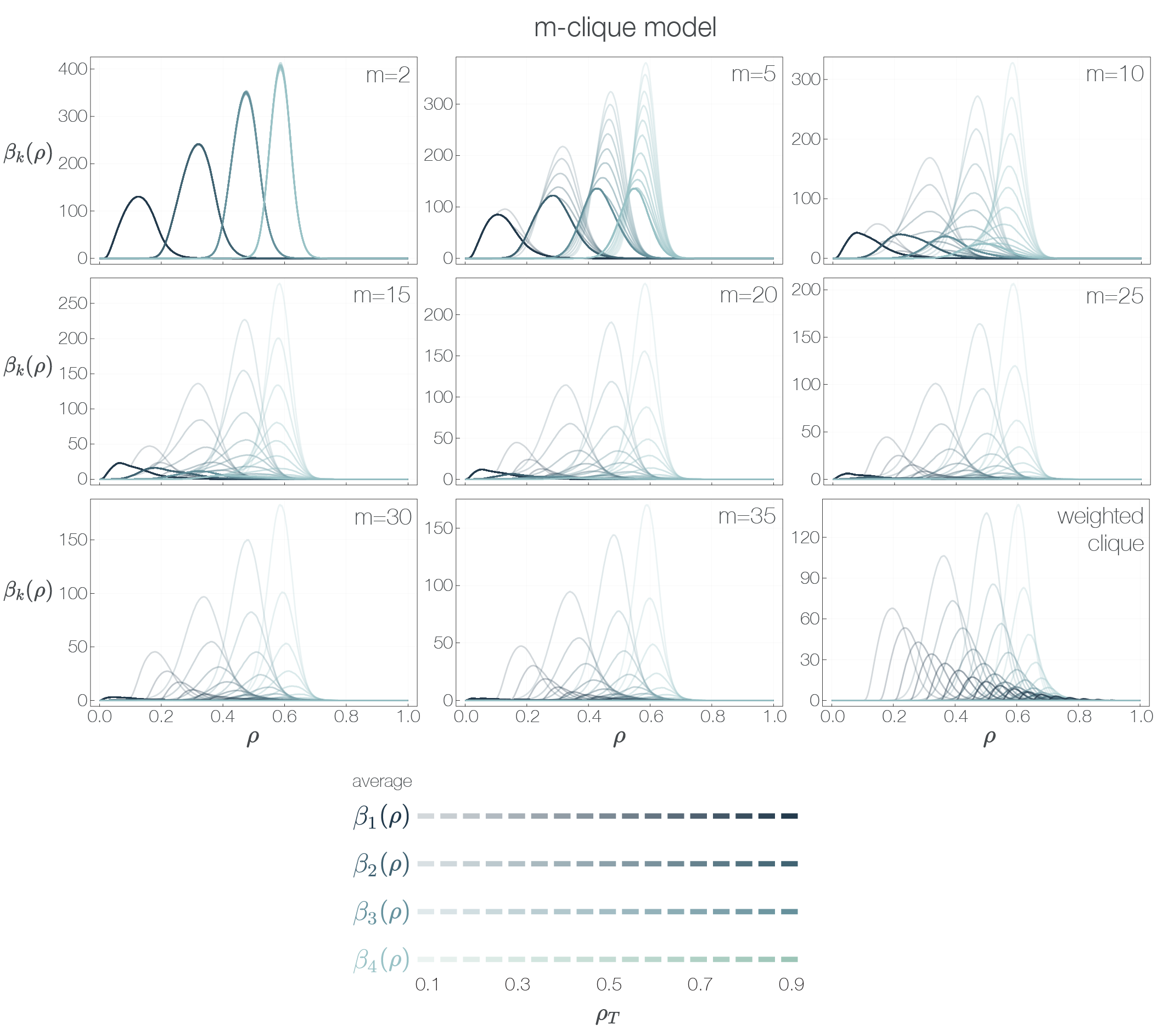}
    \caption{\textbf{Betti curves generated from the $m$-clique and weighted clique models across values of $m$, $\rho_T$.} See Fig.~\ref{fig:1} for plot details.}
    \label{sfig:clique}
\end{figure}

\clearpage
\newpage
Also in Section \ref{sec:drivers}, we introduced the $m$-clique and weighted clique models. We show the average Betti curves for all tested values of $\rho_T$ and $m$ in Fig.~\ref{sfig:clique}.

\begin{figure}[hbtp!]
    \centering
    \includegraphics{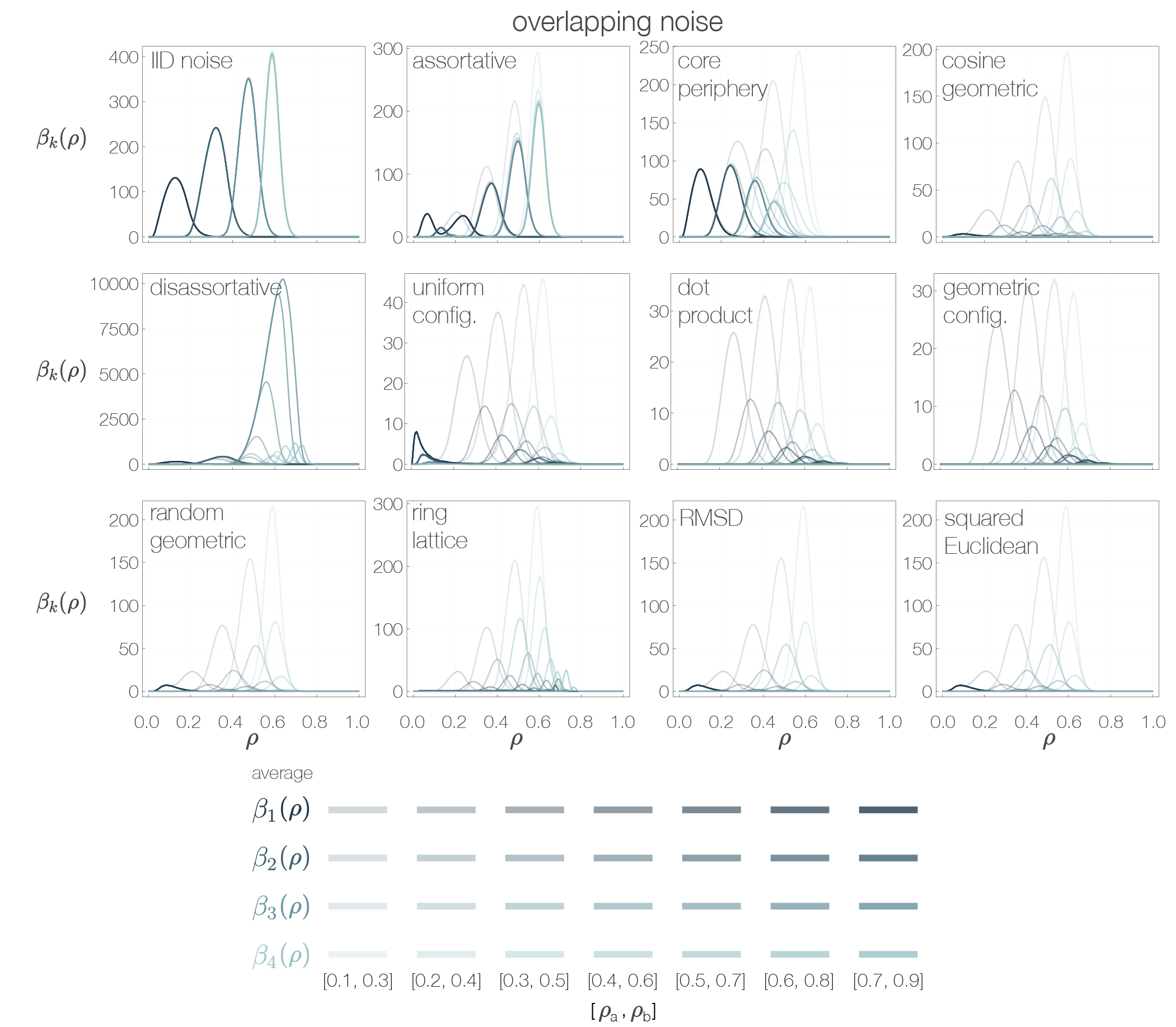}
    \caption{\textbf{Betti curves generated from adding noise to model networks in which the probability of noise increased linearly from 0 to 1 within the interval $[\rho_a, \rho_b]$.} All tested values of $[\rho_a, \rho_b]$ shown. See Fig.~\ref{fig:1} for plot details.}
    \label{sfig:overlap}
\end{figure}

\clearpage
\newpage
Finally, in Section \ref{sec:overlap} we discussed how a more realistic addition of noise to a model network could influence interpretations of Betti curves. After adding overlapping noise to all network models, we computed the persistent homology for multiple values of $\rho_a$, $\rho_b$. See Fig.~\ref{sfig:overlap} for resulting Betti curves.

\subsubsection{Classification results for all tested values of $\rho_T$}\label{sclassification}

\begin{figure}
    \centering
    \includegraphics{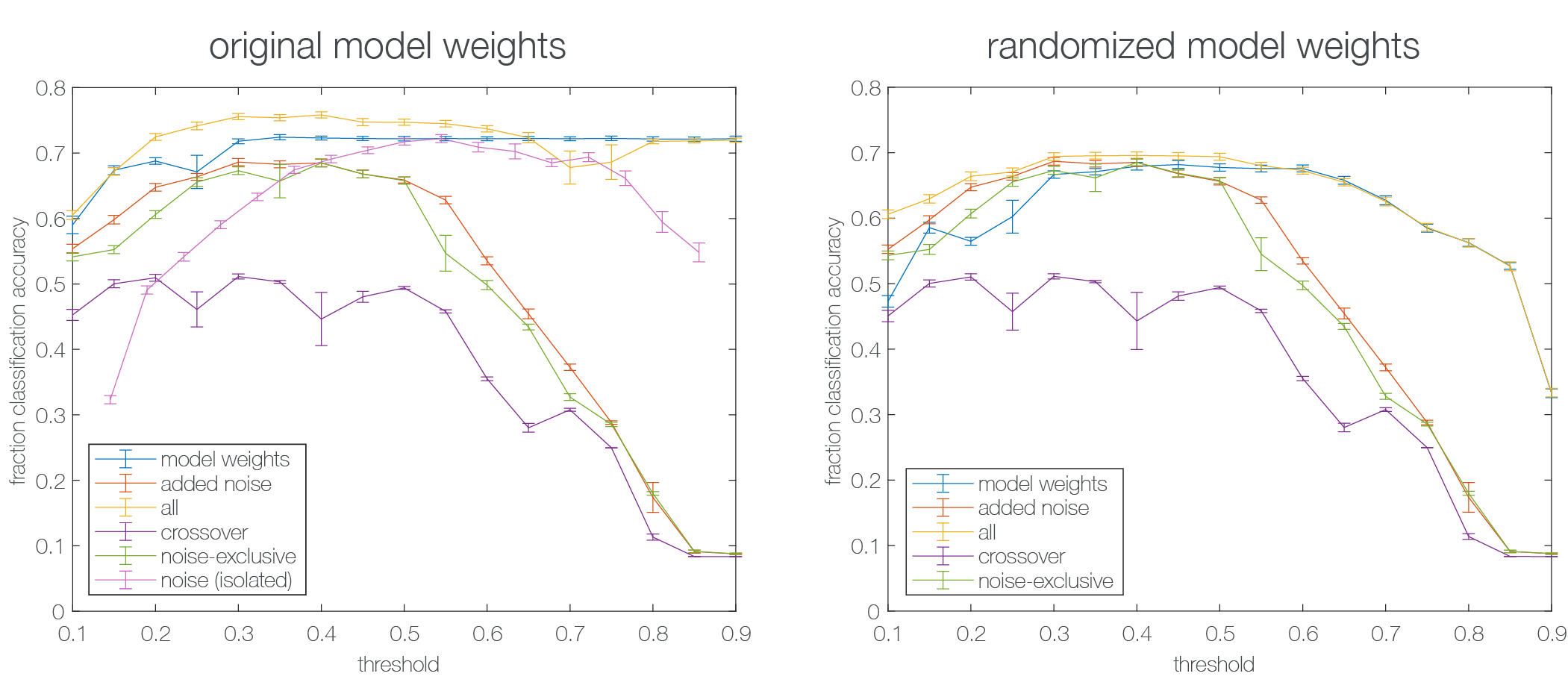}
    \caption{\textbf{Classification accuracy for experiments in the main text.} Means and standard deviations of test classification accuracy across 100 random disjoint sets of training \textit{versus} testing sets. Each point represents the mean test classification accuracy (over 100 sets) across all 12 network types when the classifier was trained on 250 graph instantiations (for a total of $12 \times 250 = 3000$ networks in the training set), and tested on a separate 250 graph instantiations (for a total of $3000$ networks in the testing set), for one threshold value and for one set of features.}
    \label{sfig:classification_accuracy}
\end{figure}

\begin{figure}
    \centering
    \includegraphics[width=6.8in]{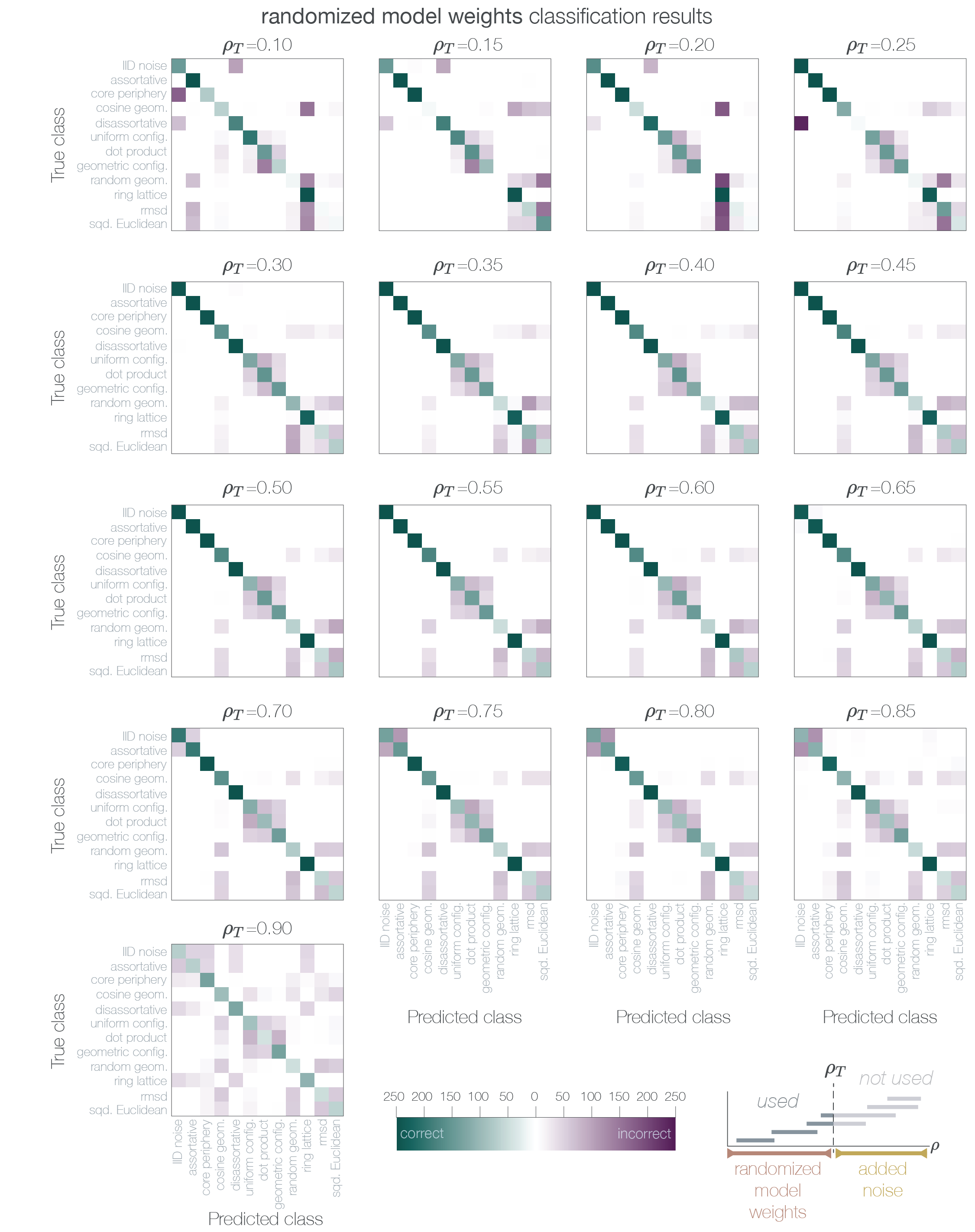}
    \caption{\textbf{Classification results for all values of $\rho_T$ using barcode summaries computed from bars in the first section of the filtration.} For these experiments, the model weights were randomized before the persistent homology was computed.}
    \label{sfig:classification_randomized}
\end{figure}

The classification experiments discussed above were performed for all values of $\rho_T$. For most experiments, confusion matrices for $\rho_T=0.5$ are shown in Fig.~\ref{fig:2}. In order to better quantify the classification performance, in Fig.~\ref{sfig:classification_accuracy} we show classification accuracy for the experiments performed with barcodes from the model weights (blue), added noise (orange), entire filtration (yellow), in addition to those performed with the crossover (purple), noise-exclusive (green), and isolated added noise (pink) barcodes (see also Fig.~\ref{fig:2}). The left plot contains results using the original model weights (original ordering), while the right plot contains results after using the randomized model weights networks (see also Fig.~\ref{sfig:classification_randomized}). Note that the added noise, crossover, and noise-exclusive results are the same in both plots since only barcodes from the added noise portion of the filtration were used in these three experiments, and the randomization of model weights would not affect this latter portion of the filtration.

\begin{figure}
    \centering
    \includegraphics[width=6.8in]{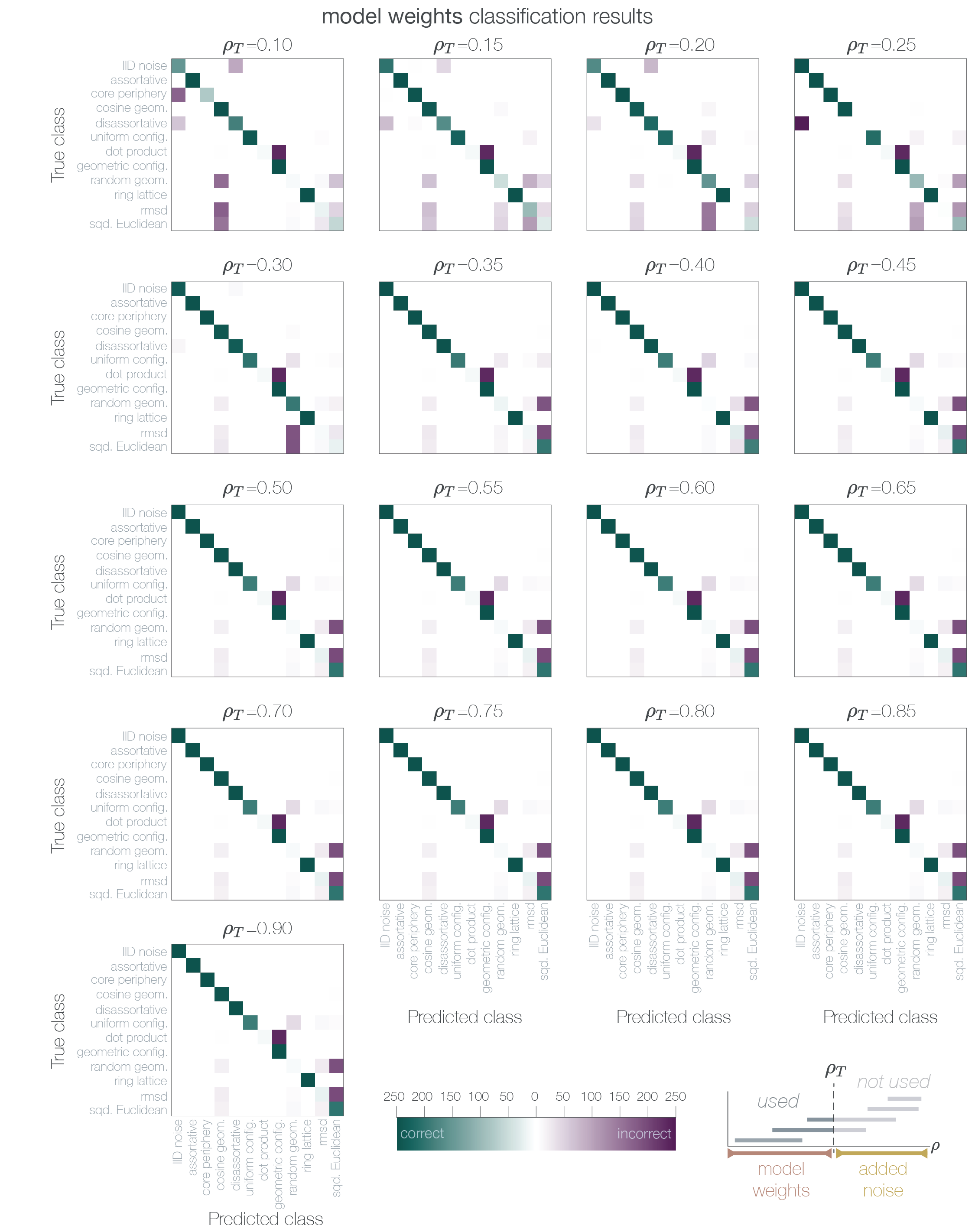}
    \caption{\textbf{Classification results for all values of $\rho_T$ using barcode summaries computed from bars in the model weights section of the filtration.}}
    \label{sfig:classification_prenoise}
\end{figure}

\begin{figure}
    \centering
    \includegraphics[width=6.8in]{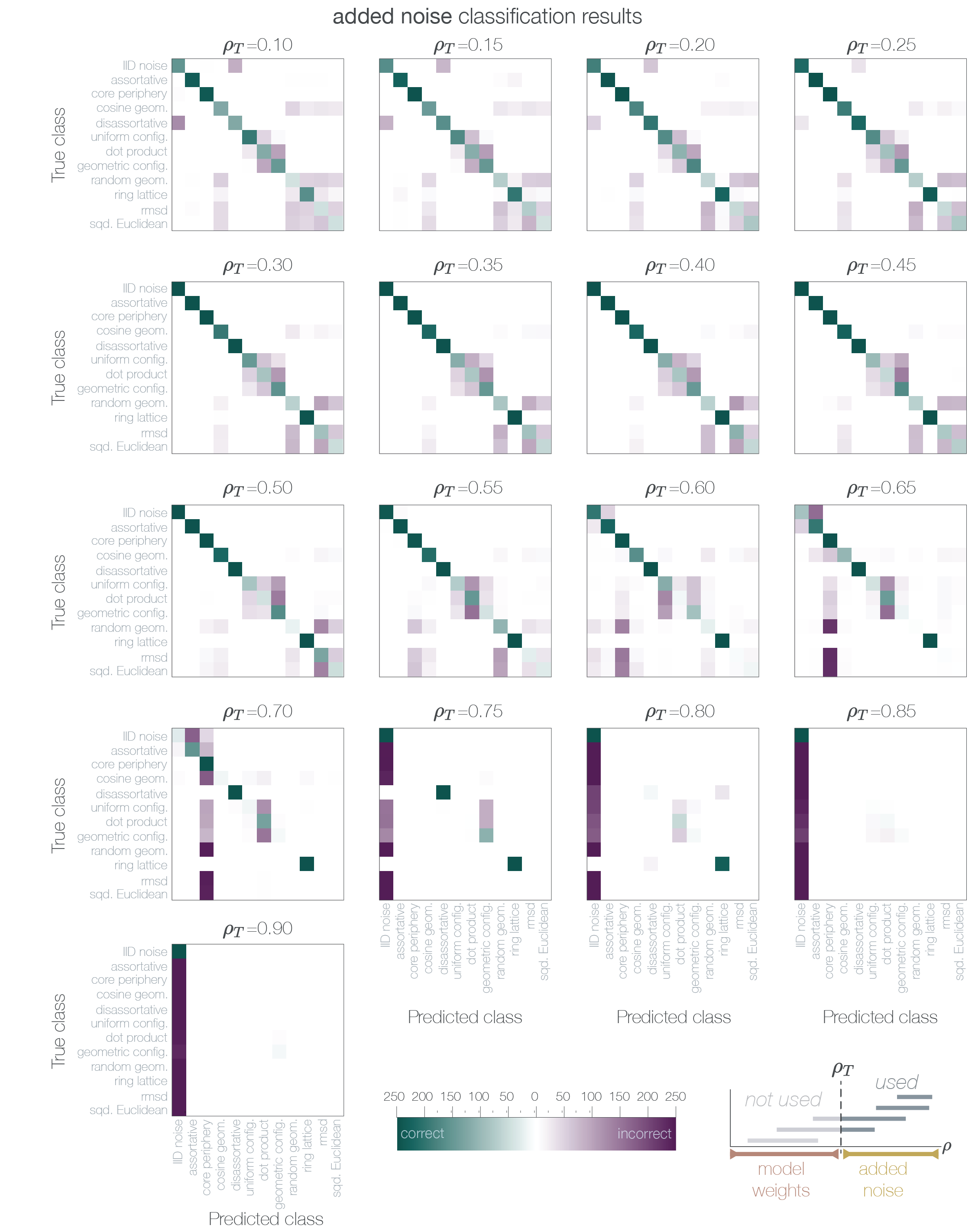}
    \caption{\textbf{Classification results for all values of $\rho_T$ using barcode summaries computed from bars in the added noise section of the filtration.}}
    \label{sfig:classification_postnoise}
\end{figure}

\begin{figure}
    \centering
    \includegraphics[width=6.8in]{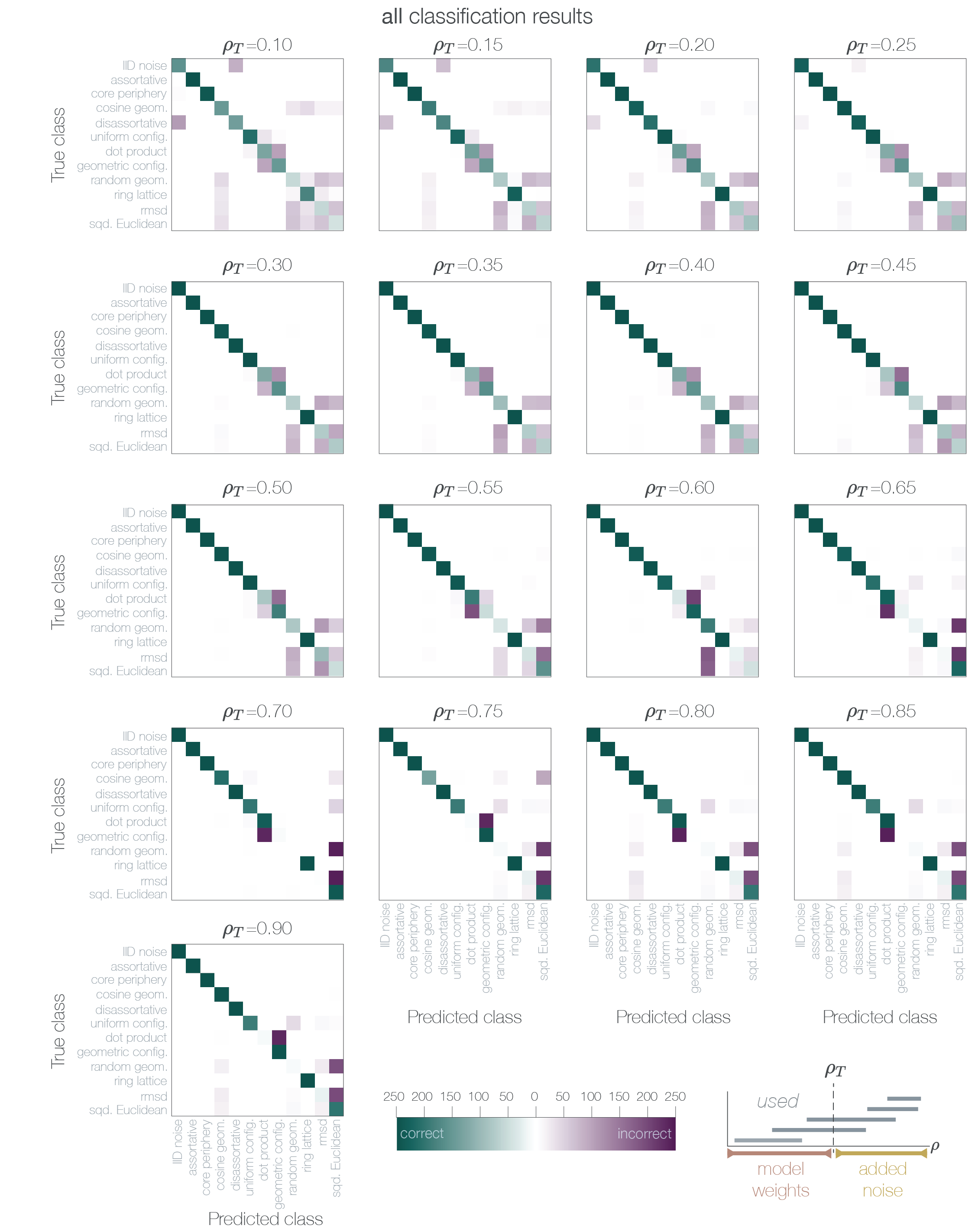}
    \caption{\textbf{Classification results for all values of $\rho_T$ using barcode summaries computed from all bars.}}
    \label{sfig:classification_all}
\end{figure}

\begin{figure}
    \centering
    \includegraphics[width=6.8in]{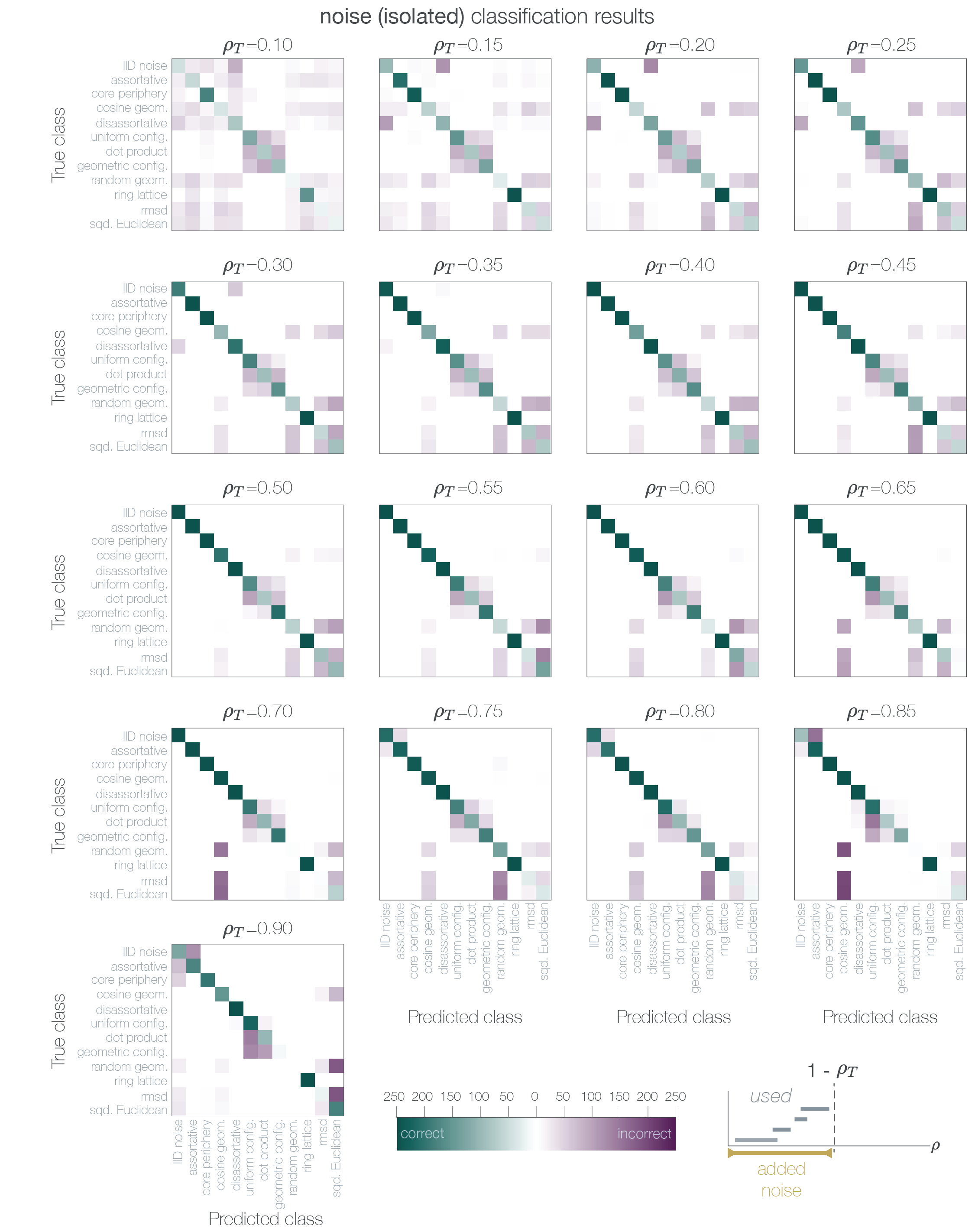}
    \caption{\textbf{Classification results for all values of $\rho_T$ using barcode summaries from the persistent homology of the added noise network in isolation.}}
    \label{sfig:classification_noiseonly}
\end{figure}

\begin{figure}
    \centering
    \includegraphics[width=6.8in]{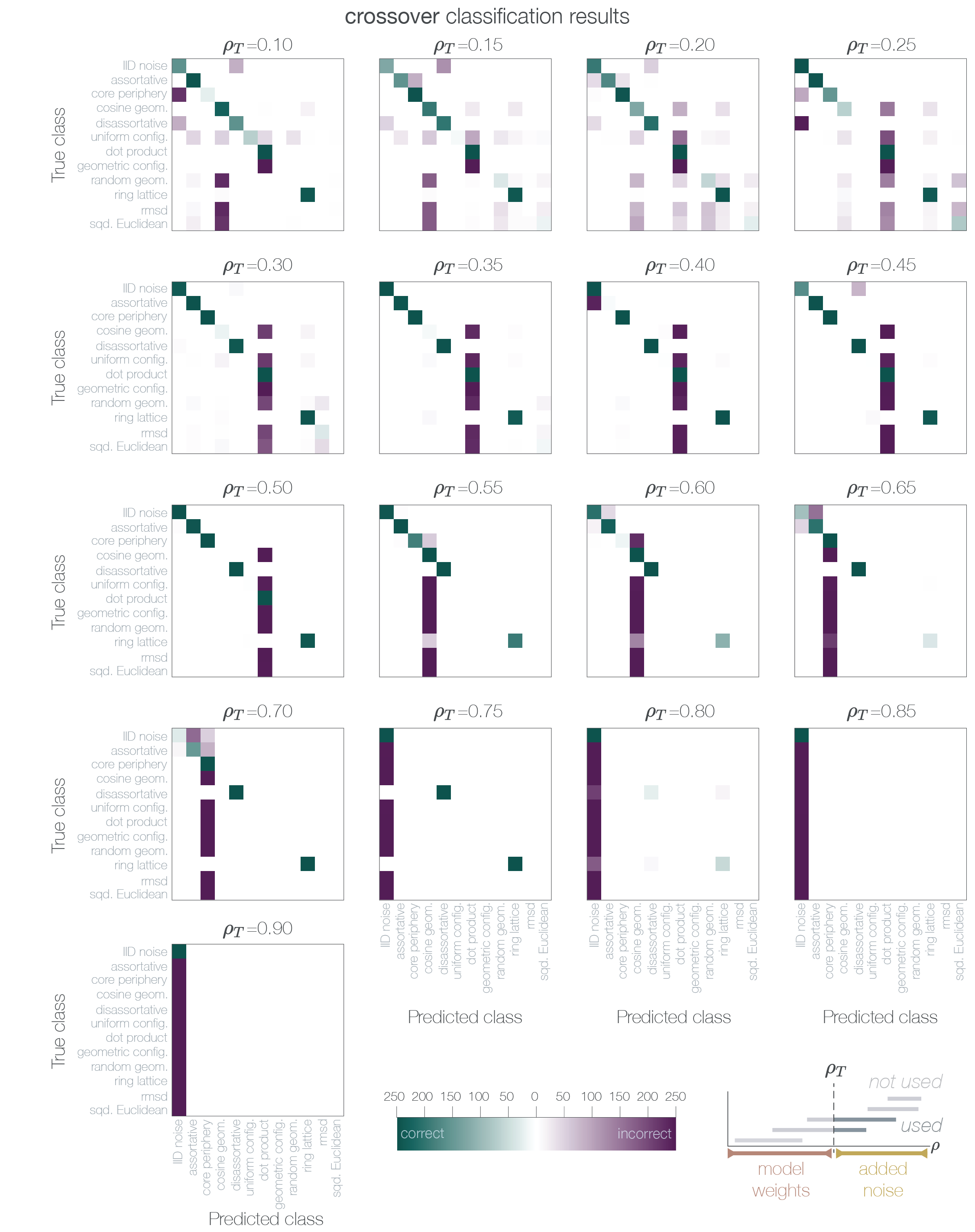}
    \caption{\textbf{Classification results for all values of $\rho_T$ using barcode summaries of those bars in the barcode that are born in the model weights section but die in the added noise section of the filtration.} All birth densities are set to $\rho_T$.}
    \label{sfig:classification_crossover}
\end{figure}

\begin{figure}
    \centering
    \includegraphics[width=6.8in]{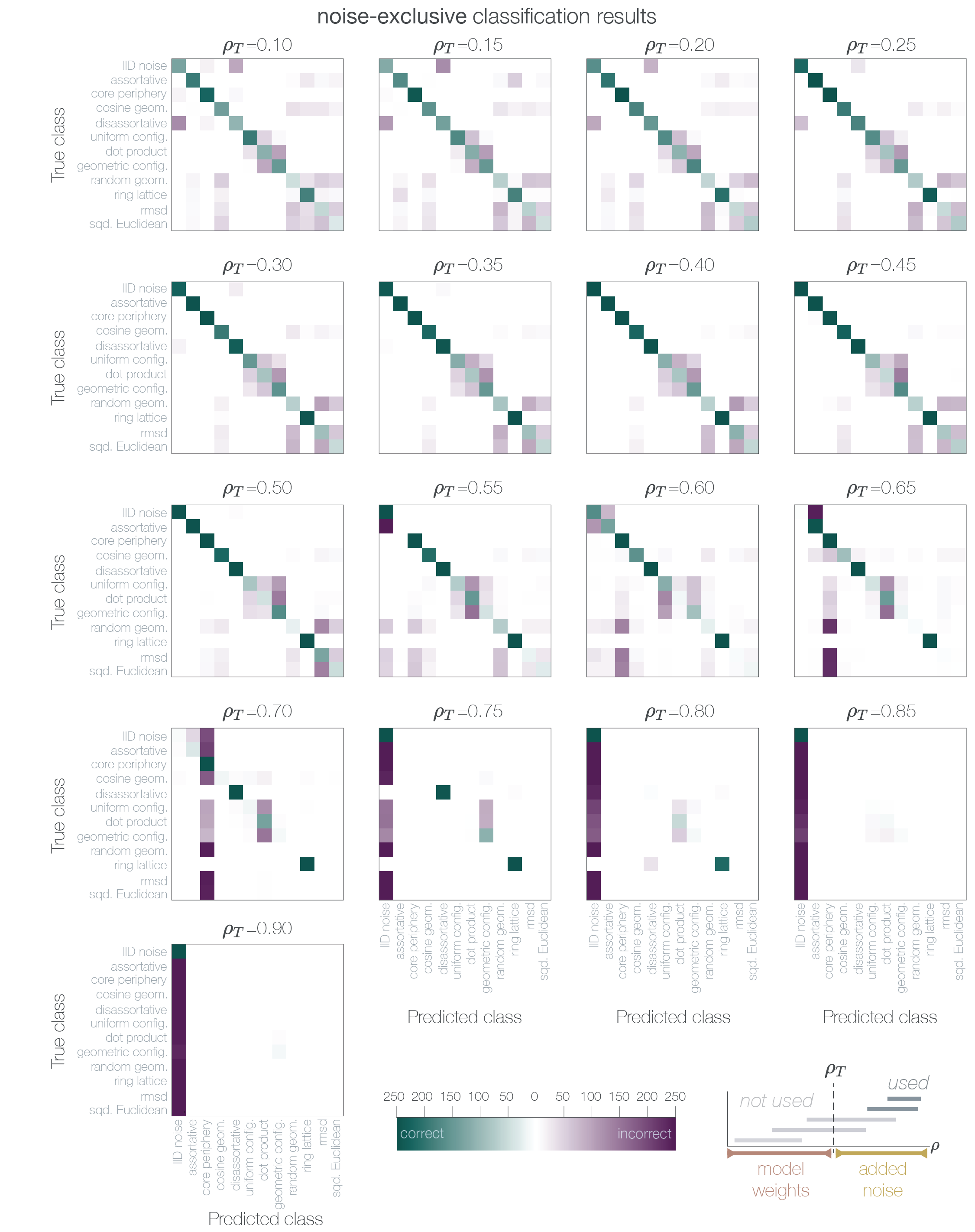}
    \caption{\textbf{Classification results for all values of $\rho_T$ using barcode summaries of those bars that are born and killed at or after $\rho_T$.}}
    \label{sfig:classification_blues}
\end{figure}

Finally, we plot confusion matrices for all values of $\rho_T$ for experiments using the model weights (Fig.~\ref{sfig:classification_prenoise}), added noise (Fig.~\ref{sfig:classification_postnoise}), the entire filtration (Fig.~\ref{sfig:classification_all}), and the isolated added noise (Fig.~\ref{sfig:classification_noiseonly}) barcodes, in addition to those using crossover (Fig.~\ref{sfig:classification_crossover}) or noise-exclusive (Fig.~\ref{sfig:classification_blues}) bars. We show the confusion matrices for the randomized model weights classification experiment in Fig.~\ref{sfig:classification_randomized}.

\subsubsection{Extended discussion of adding noise to a clique} \label{sec:adding_noise_to_clique}

In Section~\ref{sec:drivers}, we discussed a growing clique model in which at each step in the filtration we expand the one clique. Said another way, the growing graph proceeds such that if we have added exactly $\dbinom{n}{2}$ edges for some $n\leq N$, then the added edges form an $n$-clique. At these specific points in the filtration, we know that every node within the clique connects to every other node in the clique, and no other edges exist within the graph (all other nodes are isolated). This rigidity offers an opportunity to mathematically determine how noise added to the network at this point will evolve. In what follows, we will offer a brief beginning to this venture.

Suppose that we have a clique with $L < N$ nodes, and all other $M = N-L$ nodes are isolated. If we create the clique complex from this clique, we have one maximal $(L-1)$-simplex and $M$ maximal 0-simplices. Denote the number of $k$-simplices ($(k+1)$-cliques) in a binary graph by $f_k$. From Morse theory we know that
$$-f_{k-1} + f_k - f_{k+1} \leq \beta_k \leq f_k$$
where $\beta_k$ is the $k$-th Betti number. Therefore, determining expectations for $f_k$ for all $k$ will provide information about $\beta_k$ and the Euler characteristic.

Next, let us consider how $f_k$ evolves within random noise. Given a random graph on $N$ nodes in which each edge exists with probability $p<1$, $\mathbb{E}[f_k] = p^{m_k}\binom{n}{k+1}$, where $m_k = \binom{k+1}{2}$ the number of edges within a $k$-simplex. These results from the random IID graph were shown in Ref. \cite{kahle2009topology}. In contrast, in our case we have a clique on $L$ nodes and we then have $M$ isolated vertices, where $L+M = N$. So we still have work left to do.

Returning to our situation, let $p$ be the probability of each edge not in the clique $L$ existing. That is, if $p=0$, we have only the clique edges in the graph, but if $p=1$ all edges in the network will exist. Begin with $k=1$, an edge. To determine $\mathbb{E}[f_1]$ we must count the elements of three sets: (i) the number of edges within the clique (between two nodes of set $L$), (ii) the number of edges that have one node from $L$ and one from $M$, and (iii) the number of edges within the originally isolated node set (between two nodes of set $M$). Writing these out formally we have
$$ \mathbb{E}[f_1] = \binom{L}{2} + pLM + p\binom{M}{2} .$$
For 2-simplices, we also have a reasonable definition drawing from the fact that each 2-simplex has between 3 and 0 nodes in $L$ (and between 0 and 3 nodes in $M$). The edges in $L$ already exist, so we are only asking about the probability of edges forming that are not between two $L$ nodes. Again writing the above formally we have 
$$f_2 = \binom{L}{3} + p^2M\binom{L}{2} + p^3L\binom{M}{2} + p^3\binom{M}{3} .$$
Indeed any $k$-simplex will consist of between 0 and $k+1$ nodes from $L$, and consequently between $k+1$ and 0 nodes in $M$. Using this notion we can finally write the expectation for $f_k$ for such a graph with a clique of $L$ nodes and random edges added elsewhere with probability $p$ as
$$ \mathbb{E}[f_k] = \binom{L}{k+1}+ p^{m_k}\binom{M}{k+1} + \sum_{i=0}^{k-1}p^{m_k - m_i}\binom{L}{i+1}\binom{M}{k-i}.$$

Although the above calculations are far from revolutionary, they present an opportunity to derive topological features about new graph models created from combinations of simple rules.


\end{document}